\documentclass[
  prl,
  reprint,
  floatfix,
  longbibliography
  ]{revtex4-2}

\usepackage{amsmath}
\usepackage{amssymb}
\usepackage{graphicx}
\usepackage[hidelinks]{hyperref}
\usepackage{orcidlink}
\usepackage{braket}
\usepackage{bm}
\usepackage{xcolor}
\usepackage{tikz}
\usetikzlibrary{arrows.meta,calc,positioning}

\newcommand{\V}[1]{\bm{#1}}
\newcommand{\norm}[1]{\lVert #1 \rVert}
\newcommand{\smcite}{} 
\DeclareMathOperator{\tr}{tr}

\begin{document}

\title{Explicitly correlated ground states with a feedback-free variational hybrid algorithm}

\author{Benjamin Corbett\,\orcidlink{0009-0008-7108-9651}}
  \email[]{bcorbett@unm.edu}
\author{Akimasa Miyake}
  \email[]{amiyake@unm.edu}
\affiliation{Quantum New Mexico Institute, Department of Physics and Astronomy, University of New Mexico, Albuquerque, New Mexico 87106, USA}

\date{\today}

\begin{abstract}
  Explicitly correlated methods, which represent dominant short-range correlations by a diagonal Jastrow-type correlator, are designed to accurately describe systems of strongly correlated electrons, such as the fully periodic 2D Fermi--Hubbard model. However, as the reference state is compact in momentum space whereas the correlator acts naturally in real space, there is a fundamental mismatch of bases that classical algorithms resolve only by restricting the reference or the correlator. Here we propose a hybrid algorithm in which a quantum processor prepares a low-entanglement momentum-space matrix product state (MPS), precomputed classically by the transcorrelated method at the cost of a nonvariational energy, rotates it to real space, and measures in bases fixed by the Hamiltonian. Because the measurement bases are independent of the correlator, the correlator can be variationally optimized entirely in classical post-processing with no quantum-classical feedback loop. We demonstrate our algorithm on 32 qubits of Quantinuum's H2-2 with a bond-dimension-4 MPS for the half-filled $4 \times 4$ lattice and achieve a relative ground-state energy error of $16 \pm 4\%$ after postselection, reduced to $3.5 \pm 0.5\%$ by calibration using low-order moments from the MPS. Compared with other feedback-free hybrid algorithms, even under generous assumptions, our method requires fewer samples than quantum-selected configuration interaction and fewer two-qubit gates than quantum Krylov methods.
\end{abstract}

\maketitle


Explicitly correlated wave functions capture the effects of strong short-range interactions by augmenting a tractable reference state $\ket{\phi}$ with a correlator $C$ \cite{Jastrow-1955,Gutzwiller-1963,Neuscamman-Changlani-Kinder-2011}. Variational Monte Carlo (VMC) can optimize expressive diagonal correlators while preserving a variational energy \cite{Schmidt-Moskowitz-1990,Misawa-Morita-2019}, but requires repeated access to reference-state amplitudes in the basis where the correlator is diagonal. Matrix product states (MPSs), systematically improvable by the density matrix renormalization group (DMRG) \cite{White-1992,Schollwock-2011}, provide efficient amplitude access, but only in their defining basis. For the fully periodic two-dimensional Fermi--Hubbard model, momentum-space MPSs are much more accurate at a given bond dimension than real-space MPSs \cite{Xiang-1996,Ehlers-White-Noack-2017,Ehlers-Solyom-Legeza-2015,Corbett-Miyake-2025}, whereas physically motivated diagonal correlators, such as the Gutzwiller correlator, act naturally in real space \cite{Gutzwiller-1963}. Each real-space amplitude is then an overlap of the momentum-space MPS with a Slater determinant, and such overlaps are \#P-hard for arbitrary determinants \cite{Jiang-OGorman-Mahajan-2025}, obstructing direct VMC.

The transcorrelated method circumvents this basis conflict by incorporating $C$ into the similarity-transformed Hamiltonian $C^{-1}HC$ \cite{Boys-Handy-1969a,Boys-Handy-1969c}, allowing the reference state to remain in momentum space. However, increasingly expressive correlators generate increasingly complicated transcorrelated Hamiltonians \cite{Dobrautz-Luo-Alavi-2019,Baiardi-Reiher-2020,Liao-Zhai-Christlmaier-2023}; general density-density correlators have been treated on lattice models, but with the reference restricted to a single Slater determinant \cite{Wahlen-Strothman-Hoyos-2015}. Moreover, the transcorrelated energy, whether computed classically or on a quantum processor \cite{McArdle-Tew-2020,Sokolov-Dobrautz-Luo-2023,Dobrautz-Sokolov-Liao-2024}, is not variational, and physical observables generally require both left and right eigenvectors of the non-Hermitian Hamiltonian \cite{Baiardi-Reiher-2020,Dobrautz-Luo-Alavi-2019,Liao-Zhai-Christlmaier-2023}. Together, these obstacles leave the variational combination of a compact, expressive momentum-space reference state with a diagonal real-space correlator as an open problem.

Here we use a quantum processor to bridge these complementary descriptions. Transcorrelated DMRG first produces a small-bond-dimension momentum-space MPS. A quantum circuit prepares this state, performs the fermionic Fourier transformation to real space, and measures the distributions needed to evaluate both diagonal and hopping contributions to a variational explicitly correlated energy. The measurement circuits depend on the Hamiltonian terms but not on the form or parameters of the diagonal correlator. Consequently, the correlator can be selected, modified, and optimized after the quantum experiment, reusing the same measured data for every variational evaluation. Unlike conventional variational quantum algorithms \cite{Peruzzo-McClean-2014,Stanisic-Bosse-2022}, our algorithm fixes all quantum circuits before execution, with no quantum--classical optimization loop \cite{McClean-Boixo-2018,Wang-Fontana-Cerezo-2021}.

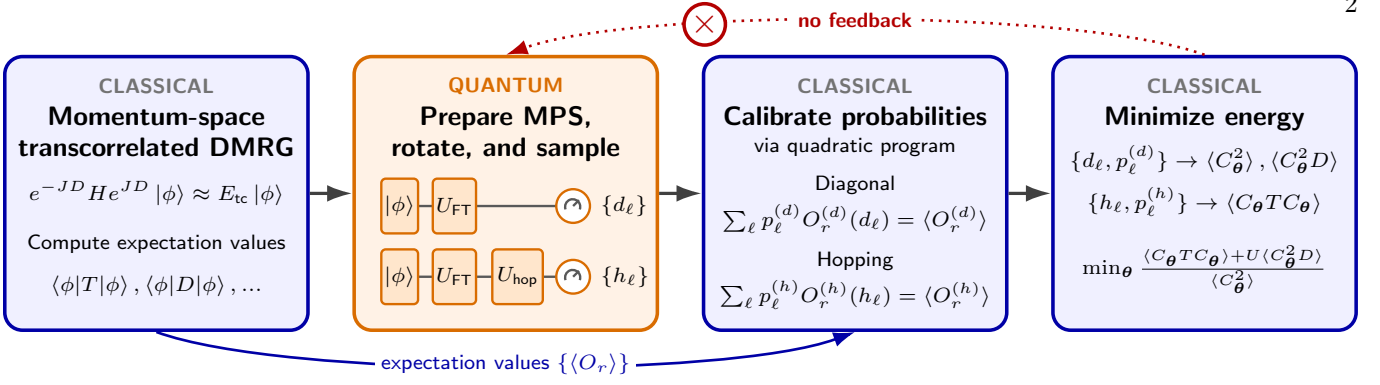
\begin{figure*}
  \vspace{-0.7cm}
  \centering
\resizebox{\textwidth}{!}{%
\begin{tikzpicture}[
    font=\sffamily,
    stage/.style={
      rounded corners=6pt,
      very thick,
      align=center,
      text width=3.25cm,
      minimum height=3.25cm,
      inner sep=5pt
    },
    classical/.style={
      stage,
      draw=blue!65!black,
      fill=blue!6
    },
    quantum/.style={
      stage,
      draw=orange!85!black,
      fill=orange!10
    },
    mainflow/.style={
      -{Latex[length=3mm,width=2mm]},
      line width=1.1pt,
      draw=black!75
    },
    auxflow/.style={
      -{Latex[length=2.6mm,width=1.8mm]},
      line width=0.9pt,
      draw=blue!65!black
    },
    blockedflow/.style={
      dotted,
      -{Latex[length=2.6mm,width=1.8mm]},
      line width=0.9pt,
      draw=red!70!black
    },
    flowlabel/.style={
      inner sep=0pt,
      font=\scriptsize\bfseries\sffamily,
      text=black!65,
      align=center
    },
    auxlabel/.style={
      fill=white,
      inner sep=2pt,
      font=\scriptsize\sffamily,
      text=blue!65!black,
      align=center
    },
    steplabel/.style={
      font=\scriptsize\bfseries\sffamily,
      text=black!55,
      align=center
    },
    titlelabel/.style={
      font=\bfseries\small\sffamily,
      align=center
    },
    sublabel/.style={
      font=\scriptsize\sffamily,
      align=center
    },
    tensor/.style={
      circle,
      draw=blue!65!black,
      fill=blue!18,
      line width=0.9pt,
      minimum size=4.8mm,
      inner sep=0pt
    },
    gate/.style={
      rounded corners=1.5pt,
      draw=orange!85!black,
      fill=orange!20,
      line width=0.8pt,
      minimum width=5.5mm,
      inner sep=1pt
    },
    meter/.style={
      circle,
      draw=orange!85!black,
      fill=white,
      line width=0.8pt,
      minimum size=4.0mm,
      inner sep=0pt
    },
    iconline/.style={
      line width=0.8pt,
      draw=black!65
    },
    bar/.style={
      draw=blue!65!black,
      fill=blue!20,
      line width=0.5pt
    },
    calibratedbar/.style={
      bar,
      fill=blue!38
    },
    histaxis/.style={
      line width=0.55pt,
      draw=black!55
    },
    energycurve/.style={
      line width=1.0pt,
      draw=blue!65!black
    },
    minimumarrow/.style={
      -{Latex[length=2.8mm,width=2.0mm]},
      line width=1.4pt,
      draw=blue!65!black
    },
    calibrateflow/.style={
      -{Latex[length=2.1mm,width=1.4mm]},
      line width=0.8pt,
      draw=blue!65!black
    },
    cross/.style={
      circle,
      solid,
      draw=red!70!black,
      fill=white,
      text=red!70!black,
      very thick,
      font=\large\bfseries,
      inner sep=0.6pt
    }
  ]

  \node[classical] (dmrg) at (0,0) {};
  \node[quantum]   (qpu)  at (4.15,0) {};
  \node[classical] (fit)  at (8.30,0) {};
  \node[classical] (vmc)  at (12.45,0) {};

  \node[steplabel] at ([yshift=1.28cm]dmrg.center) {CLASSICAL};
  \node[titlelabel, anchor=north]
    at ([yshift=-0.50cm]dmrg.north)
    {Momentum-space \\ transcorrelated DMRG};
  \node[sublabel] at ([yshift=-0.0cm]dmrg.center) {$e^{-J D} H e^{J D} \ket{\phi} \approx E_\text{tc} \ket{\phi}$};
  \node[sublabel] at ([yshift=-0.6cm]dmrg.center) {Compute expectation values};
  \node[sublabel] at ([yshift=-1.1cm]dmrg.center) {$\braket{\phi | T | \phi}, \braket{\phi | D | \phi}, ...$};

  \node[steplabel, text=orange!85!black] at ([yshift=1.28cm]qpu.center) {QUANTUM};
  \node[titlelabel, anchor=north]
    at ([yshift=-0.50cm]qpu.north)
    {Prepare MPS, \\ rotate, and sample};
  \begin{scope}[shift={($(qpu.center)+(0.34,-0.57)$)}]
    \draw[iconline] (-1.5,0.42) -- (0.42,0.42);
    \node[gate, minimum width=4mm, minimum height=6.5mm, font=\scriptsize\sffamily]
      at (-1.6,0.42) {$\ket{\phi}$};
    \node[gate, minimum width=4mm, minimum height=6.5mm, font=\scriptsize\sffamily]
      at (-0.95,0.42) {$U_\text{FT}$};
    \node[meter] (meterdiag) at (0.46,0.42) {};
    \draw[iconline] ($(meterdiag.center)+(-0.09,-0.04)$)
      arc[start angle=200,end angle=-20,radius=0.09];
    \draw[iconline] (meterdiag.center) -- ++(0.07,0.07);
    \node[anchor=west, font=\scriptsize\sffamily]
      at (meterdiag.east) {$\{d_\ell\}$};

    \draw[iconline] (-1.5,-0.42) -- (0.42,-0.42);
    \node[gate, minimum width=4mm, minimum height=6.5mm, font=\scriptsize\sffamily]
      at (-1.6,-0.42) {$\ket{\phi}$};
    \node[gate, minimum width=4mm, minimum height=6.5mm, font=\scriptsize\sffamily]
      at (-0.95,-0.42) {$U_\text{FT}$};
    \node[gate, minimum width=6mm, minimum height=6.5mm, font=\scriptsize\sffamily]
      at (-0.2,-0.42) {$U_\text{hop}$};
    \node[meter] (meterhop) at (0.46,-0.42) {};
    \draw[iconline] ($(meterhop.center)+(-0.09,-0.04)$)
      arc[start angle=200,end angle=-20,radius=0.09];
    \draw[iconline] (meterhop.center) -- ++(0.07,0.07);
    \node[anchor=west, font=\scriptsize\sffamily]
      at (meterhop.east) {$\{h_\ell\}$};
  \end{scope}

  \node[steplabel] at ([yshift=1.28cm]fit.center) {CLASSICAL};
  \node[titlelabel, anchor=north] at ([yshift=-0.50cm]fit.north) {Calibrate probabilities};
  \node[sublabel] at ([yshift=0.55cm]fit.center) {via quadratic program};
  \node[sublabel] at ([yshift=0.1cm]fit.center) {Diagonal};
  \node[sublabel] at ([yshift=-0.3cm]fit.center) {$\sum_\ell p^{(d)}_\ell O^{(d)}_r(d_\ell) = \braket{O^{(d)}_r}$};
  \node[sublabel] at ([yshift=-0.8cm]fit.center) {Hopping};
  \node[sublabel] at ([yshift=-1.2cm]fit.center) {$\sum_\ell p^{(h)}_\ell O^{(h)}_r(h_\ell) = \braket{O^{(h)}_r}$};

  \node[steplabel] at ([yshift=1.28cm]vmc.center) {CLASSICAL};
  \node[titlelabel, anchor=north] at ([yshift=-0.50cm]vmc.north) {Minimize energy};
  \node[sublabel] at ([yshift=0.4cm]vmc.center) {$\{ d_\ell, p^{(d)}_\ell\} \rightarrow \braket{C^2_{\V{\theta}}}, \braket{C^2_{\V{\theta}} D}$};
  \node[sublabel] at ([yshift=-0.1cm]vmc.center) {$\{ h_\ell, p^{(h)}_\ell\} \rightarrow \braket{C_{\V{\theta}}T C_{\V{\theta}}}$};
  \node[sublabel] at ([yshift=-0.9cm]vmc.center) {$\min_{\V{\theta}} \frac{\braket{C_{\V{\theta}} T C_{\V{\theta}}} + U \braket{C^2_{\V{\theta}} D}}{\braket{C^2_{\V{\theta}}}}$};

  \draw[mainflow]
    (dmrg.east) -- (qpu.west);
  \draw[mainflow]
    (qpu.east) -- (fit.west);
  \draw[mainflow]
    (fit.east) -- (vmc.west);

  \draw[auxflow]
    (dmrg.south) to[out=-20,in=-160, looseness=0.5]
    node[auxlabel, yshift=1pt] {expectation values $\{ \braket{O_r} \}$}
    (fit.south);

  \draw[blockedflow]
    (vmc.north) to[out=160,in=20, looseness=0.5]
    node[cross, pos=0.68] {$\times$}
    node[
      flowlabel,
      midway,
      fill=white,
      inner sep=2pt,
      text=red!70!black,
      font=\scriptsize\bfseries\sffamily
    ] {no feedback}
    (qpu.north);
\end{tikzpicture}%
}
  \vspace{-0.7cm}
  \caption{\label{fig:algorithm-overview}
    Transcorrelated momentum-space DMRG with the Gutzwiller factor $e^{J D}$ produces the approximate ground state $\ket{\phi}$ and nonvariational energy $E_\text{tc}$. A quantum processor prepares $\ket{\phi}$, rotates it to real space, and measures in fixed diagonal and hopping bases. The samples are calibrated to reproduce known MPS expectation values and reused to optimize arbitrary real-space diagonal correlators $C_{\V{\theta}}$ entirely in classical post-processing, without a quantum--classical feedback loop.
  }
\end{figure*}

\section{Quantum-classical hybrid algorithm}

We consider the Fermi--Hubbard Hamiltonian \cite{Hubbard-1963} on square lattices with uniform nearest-neighbor hopping and periodic boundary conditions,
\begin{equation*}
  H = -t \sum_{\braket{\V{x}, \V{y}}} \sum_{\sigma \in \{\uparrow, \downarrow\}} a^\dagger_{\V{x} \sigma} a_{\V{y} \sigma} + U D \ , \quad D = \sum_{\V{x}} n_{\V{x} \uparrow} n_{\V{x} \downarrow} \ ,
\end{equation*}
where $\braket{\V{x}, \V{y}}$ runs over all ordered pairs of neighboring sites, and $n_{\V{x} \sigma} = a^\dagger_{\V{x} \sigma} a_{\V{x} \sigma}$.

We target the variationally correlated state $\ket{\psi(\V{\theta})} = C_{\V{\theta}} \ket{\phi}$, where
\begin{equation} \label{eq:explicitly-correlated-energy}
  C_{\V{\theta}} = \exp \left( \sum_r \theta_r D_r \right) \ , \quad E_\text{ec}(\V{\theta}) = \frac{\braket{\phi | C_{\V{\theta}} H C_{\V{\theta}} | \phi}}{\braket{\phi | C_{\V{\theta}}^2 | \phi}} \ .
\end{equation}
The real parameters $\theta_r$ multiply Hermitian operators $D_r$ diagonal in the real-space occupation basis.

\subsection{Momentum-space transcorrelated DMRG}

The algorithm (Fig.~\ref{fig:algorithm-overview}) begins with a normalized momentum-space MPS $\ket{\phi}$ from transcorrelated DMRG, which uses the Gutzwiller factor $C_\text{tc} = e^{J D}$ with $J < 0$, optimizes $J$ and the MPS self-consistently by the residual-based method of Ref.~\cite{Corbett-Miyake-2025} (End Matter), and yields the nonvariational energy $E_\text{tc} = \braket{\phi | C_\text{tc}^{-1} H C_\text{tc} | \phi}$. Although $C_\text{tc}$ has the same form as the one-parameter Gutzwiller case of Eq.~\eqref{eq:explicitly-correlated-energy}, $C_\theta = e^{\theta D}$, $J$ defines the similarity transformation that generates $\ket{\phi}$, whereas $\theta$ variationally correlates that fixed reference in classical post-processing.

\subsection{Prepare MPS, rotate, and sample}

A quantum processor prepares the MPS $\ket{\phi}$ \cite{Ran-2020,Rudolph-Chen-2023} and rotates it to real space with the fermionic Fourier transformation $U_\text{FT}$, ${a_{\V{x} \sigma} = N^{-1/2} \sum_{\V{k}} e^{-i \V{k} \cdot \V{x}} a_{\V{k} \sigma}}$ \cite{Jiang-Sung-Kechedzhi-2018,Kivlichan-McClean-2018}, where $N$ is the number of sites.

The upper circuit in the second box of Fig.~\ref{fig:algorithm-overview} measures this state in the occupation basis. Repeating this circuit $M^{(d)}$ times produces outcomes $\{ d_\ell \}_\ell$, where $d_\ell \in \{ 0, 1 \}^{2 N}$, drawn from $\Pr(d_\ell) = \left| \braket{d_\ell | U_\text{FT} | \phi} \right|^2$. Because the correlator $C$ and $D$ are diagonal in this basis, each outcome determines $C(d_\ell)$ and $D(d_\ell)$. Assigning weight $p^{(d)}_\ell$ to each sample gives
\begin{equation} \label{eq:diagonal-estimators}
  \braket{C^2} \simeq \sum_{ \ell = 1}^{M^{(d)}} p^{(d)}_\ell C(d_\ell)^2 \ , \braket{C^2 D} \simeq \sum_{\ell = 1}^{M^{(d)}} p^{(d)}_\ell C(d_\ell)^2 D(d_\ell) \ .
\end{equation}

For each hopping pair $s = (j, k)$, where a single index labels a spin orbital, let $T_s = -t (a_j^\dagger a_k + a_k^\dagger a_j)$. After $U_\text{FT}$, the lower circuit applies a number-conserving matchgate $U_\text{hop}^{(s)}$ that diagonalizes $T_s$. Repeating this circuit $M^{(s)}$ times produces outcomes $\{ h_\ell^{(s)} \}_\ell$ drawn from $\Pr_s(h) = |\braket{h | U_\text{hop}^{(s)} U_\text{FT} | \phi}|^2$. Each outcome determines both $T_s(h_\ell^{(s)})$, which includes the Jordan--Wigner sign \cite{Jordan-Wigner-1928}, and the occupation of every spin orbital other than $j$ and $k$. Let $C(h^{(s)}_{\ell} \,|\, 1_j, 0_k)$ and $C(h^{(s)}_{\ell} \,|\, 0_j, 1_k)$ be the correlator evaluated using the occupations specified by $h^{(s)}_\ell$, with the occupations at $j$ and $k$ fixed as shown. Assigning weight $p_\ell^{(s)}$ to each sample gives
\begin{equation} \label{eq:hopping-estimator}
  \braket{C T_s C} \simeq \sum_{\ell = 1}^{M^{(s)}} p^{(s)}_\ell C(h^{(s)}_\ell \,|\, 1_j, 0_k) C(h^{(s)}_\ell \,|\, 0_j, 1_k) T_s(h^{(s)}_\ell) \ ,
\end{equation}
which we then sum over $s$ to obtain $\braket{C T C}$. A derivation of Eq.~\eqref{eq:hopping-estimator} is given in the End Matter. Before calibration, the sample weights are $p^{(d)}_\ell = 1/M^{(d)}$ and $p^{(s)}_\ell = 1/M^{(s)}$.

\begin{figure*}
  \centering
  \begin{minipage}[t]{0.325\textwidth}
    \makebox[\linewidth][l]{\textbf{(a)}}\par
    \includegraphics[width=\linewidth]{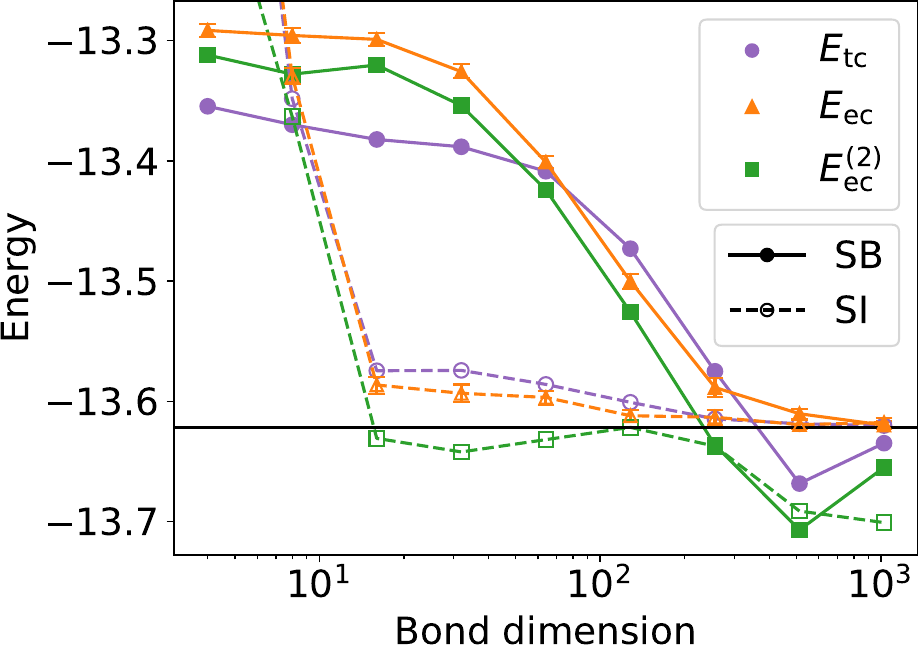}
  \end{minipage}
  \hfill
  \begin{minipage}[t]{0.325\textwidth}
    \makebox[\linewidth][l]{\textbf{(b)}}\par
    \includegraphics[width=\linewidth]{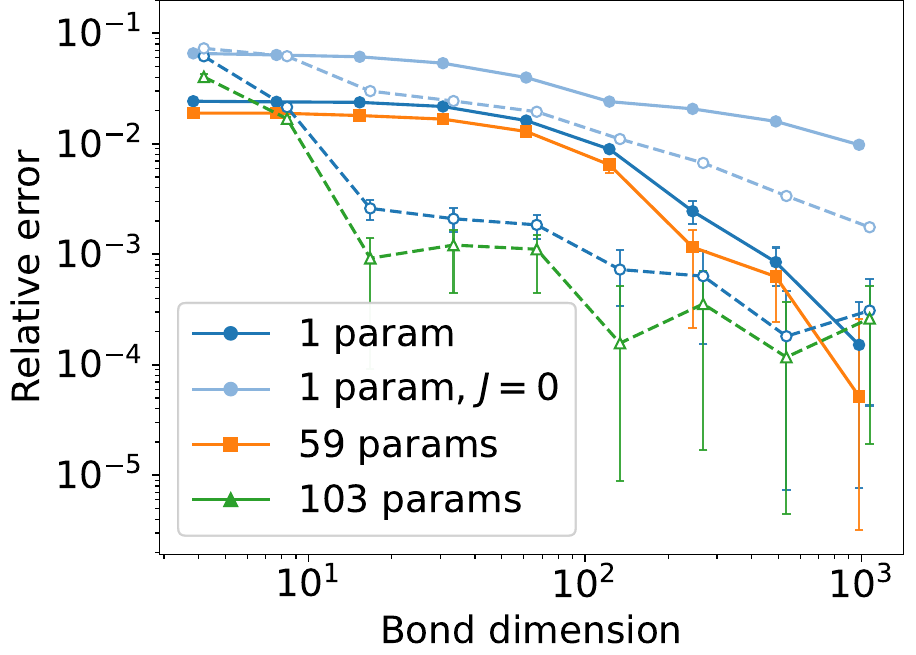}
  \end{minipage}
  \hfill
  \begin{minipage}[t]{0.325\textwidth}
    \makebox[\linewidth][l]{\textbf{(c)}}\par
    \includegraphics[width=\linewidth]{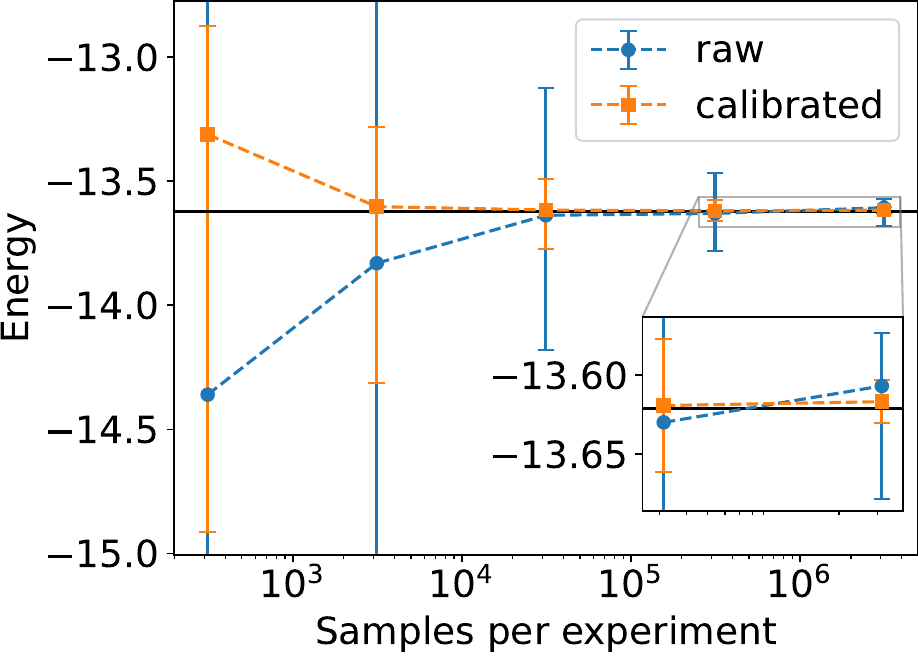}
  \end{minipage}
  \caption{\label{fig:classical-results}
    \textbf{(a)} Transcorrelated energy $E_\text{tc}$, sampled explicitly correlated energy $E_\text{ec}$, and the minimum of its second-order Taylor approximation $E^{(2)}_\text{ec}$, Eq.~\eqref{eq:ec-taylor}, for the one-parameter Gutzwiller correlator versus bond dimension $\chi$ for the spin-blocked (SB) and spin-interleaved (SI) orderings.
    \textbf{(b)} Relative $E_\text{ec}$ errors for the Gutzwiller, 59-parameter, and 103-parameter correlators and for a Gutzwiller correlator on a non-transcorrelated ($J = 0$) reference, which barely improves on direct DMRG (not shown).
    \textbf{(c)} Median and central 68\% interval from partitioning 157 million samples for the $\chi = 1024$ SI MPS into experiments of varying size, with and without calibration.
    Horizontal lines mark the exact ground-state energy $-13.621855$ in (a) and the infinite-sample energy $-13.621361$ in (c).
  }
\end{figure*}

\subsection{Calibrate probabilities}

In the infinite-sample limit, defined by exact state preparation and noiseless measurements with $M^{(d)}, M^{(s)} \to \infty$, Eqs.~\eqref{eq:diagonal-estimators} and \eqref{eq:hopping-estimator} evaluate $E_\text{ec}(\V{\theta})$ exactly for the prepared reference and any post-processed correlator. Finite sampling and hardware errors perturb the measured distributions; we reweight them to reproduce reference-state expectations known from the classical momentum-space MPS, such as $\braket{\phi | D | \phi}$, reducing both errors.

Since we target the explicitly correlated energy in Eq.~\eqref{eq:explicitly-correlated-energy}, natural constraints appear in its second-order Taylor approximation about $\V{\theta} = 0$,
\begin{align} \label{eq:ec-taylor}
  E^{(2)}_\text{ec}(\V{\theta}) &= \braket{H} + \sum_r \braket{ \{ \delta H, \delta D_r \}} \theta_r \nonumber \\
  &+ \frac{1}{2} \sum_{rs} \braket{ \{ \{ \delta H, \delta D_r \}, \delta D_s \} } \theta_r \theta_s \ ,
\end{align}
where $\delta A = A - \braket{A}$ and $\braket{A} = \braket{\phi | A | \phi}$. Calibrating diagonal samples with $\braket{D}$, $\braket{D_r}$, and $\braket{D D_r}$, and hopping samples with $\braket{T}$, $\Re \braket{T D_r}$, and $\Re \braket{T D_r D_s + D_r T D_s}$, fixes the zeroth- and first-order terms and the second-order hopping contribution. These constraints pin only low-order moments of the sampled distributions; the calibrated samples still evaluate $E_\text{ec}(\V{\theta})$ without expansion (Fig.~\ref{fig:energy-vs-j}). We exclude $\braket{D D_r D_s}$ because its evaluation can be more expensive than transcorrelated DMRG itself, and is not always viable. An equality-constrained quadratic program finds the minimum-norm multiplicative change to the empirical frequencies that reproduces these targets, rejecting any solution containing a negative probability (End Matter).

\subsection{Minimize energy}

Using the calibrated weights in Eqs.~\eqref{eq:diagonal-estimators} and \eqref{eq:hopping-estimator}, we evaluate the variational energy in Eq.~\eqref{eq:explicitly-correlated-energy} and minimize it over $\V{\theta}$ entirely in classical post-processing.

\section{General theory and demonstration for the 2D Fermi--Hubbard model}

We simulate the half-filled $4 \times 4$ model at $U / t = 4$, a regime with established benchmarks \cite{Shi-Zhang-2013,LeBlanc-Antipov-2015,Qin-Shi-Zhang-2016}, and report energies in units of $t$. At this size the real-space rotation can still be performed exactly on a classical computer, so every stage of the algorithm can be benchmarked; by $10 \times 10$, even a truncated classical rotation of a $\chi = 16$ MPS retains a 4\% energy error after more than two days of computation, as shown in the Supplemental Material (SM)\smcite.

Figure~\ref{fig:classical-results}(a) compares the transcorrelated energy with the explicitly correlated energy obtained using the one-parameter Gutzwiller correlator, for MPSs in the spin-blocked (SB) and spin-interleaved (SI) qubit orderings (End Matter). For the SB ordering, the transcorrelated energy falls below the ground-state energy at $\chi = 512$ and 1024, illustrating its nonvariational nature. By contrast, the explicitly correlated energy converges toward the ground-state energy and remains statistically consistent with the variational bound. We also plot the minimum of the quadratic approximation $E^{(2)}_\text{ec}(\theta)$ defined in Eq.~\eqref{eq:ec-taylor}. This approximation falls below the exact ground-state energy and differs substantially from $E_\text{ec}$ at large bond dimensions. Thus, the low-order expectation values used for calibration are insufficient to reproduce the optimized variational energy. Additionally, the approximation can break down with more parameters, since the quadratic form need not possess a local minimum.

Figure~\ref{fig:classical-results}(b) compares the relative energy error, $|E_\text{ec} - E_\text{gs}| / |E_\text{gs}|$, for the Gutzwiller correlator and the 59-parameter SB and 103-parameter SI correlators defined in the SM\smcite. For each reference MPS, all correlators use the same samples, calibrated only with Gutzwiller constraints. The improvement is statistically resolved for $\chi \le 128$ and is largest at small bond dimension, the regime relevant to hardware state preparation. At $\chi = 16$, the 103-parameter SI correlator gives $-13.609 \pm 0.005$. Achieving an energy of $-13.6$ requires $\chi = 1024$ with non-transcorrelated momentum-space DMRG or $\chi = 4096$ in real space. Counting symmetry-allowed nonzero MPS tensor entries, our ansatz has 1012 parameters, a 141-fold reduction over the momentum-space MPS.

Figure~\ref{fig:classical-results}(c) isolates the effect of calibration by partitioning the 157 million samples from the $\chi = 1024$ SI MPS into experiments of increasing size. Calibration narrows the central 68\% interval by factors of 3.3--4.4. At small sample sizes, optimization produces long lower tails by exploiting statistical inconsistencies among the sampled distributions, but among successful experiments we resolve no calibration-induced bias at any size considered.

\begin{figure}
  \includegraphics[width=\columnwidth]{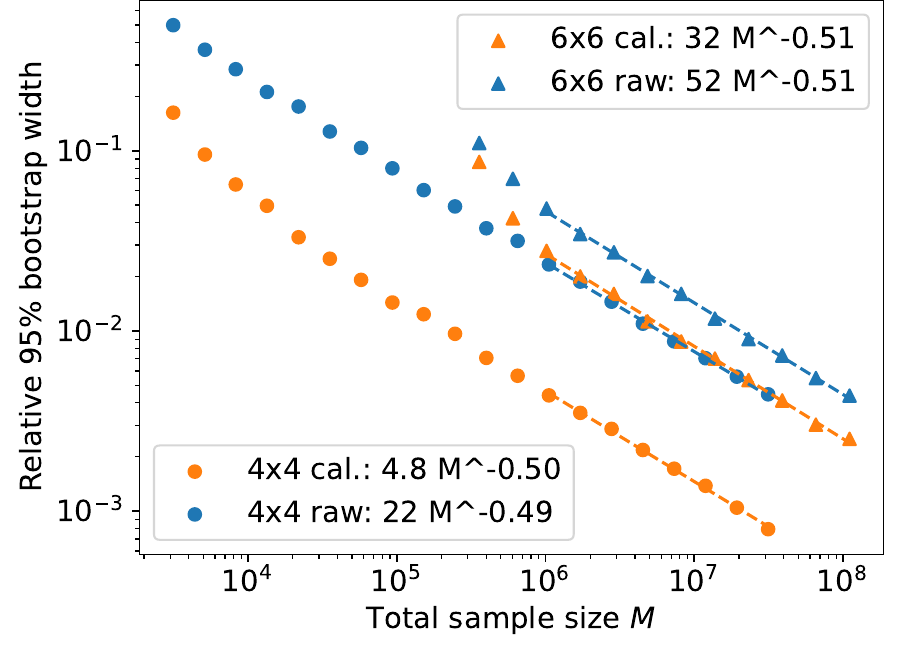}
  \caption{\label{fig:ci-width-vs-samples}
    Relative width of 95\% bootstrap confidence intervals versus total samples $M = M^{(d)} + 4 N M^{(s)}$ for raw and calibrated (cal.) energies from $\chi = 4$ SB MPSs with the Gutzwiller correlator. Widths are divided by the reference ground-state energy magnitude, from exact diagonalization for $4 \times 4$ and auxiliary-field QMC for $6 \times 6$ \cite{Qin-Shi-Zhang-2016}. The allocation is $M^{(d)} : M^{(s)} = 122 : 3$. Plotted budgets have at least 95\% successful replicates; dashed power-law fits use $M \geq 10^6$.
  }
\end{figure}
To investigate sampling cost versus lattice size $N$ (see SM\smcite), we collected 110.8 million samples from a $\chi = 4$ SB MPS for the $6 \times 6$ system after a truncated real-space rotation with a 0.02\% energy error. Figure~\ref{fig:ci-width-vs-samples} shows the relative 95\% bootstrap widths before and after calibration, following the expected $1 / \sqrt{M}$ behavior. At fixed relative precision, the raw $6 \times 6$ data require about $3.4 \times$ more samples than the $4 \times 4$ data. The importance-sampling factor $R = \braket{\phi | C_{\V{\theta}}^4 | \phi} / \braket{\phi | C_{\V{\theta}}^2 | \phi}^2$ sets the sample count up to polynomial factors \cite{Agapiou-Papaspiliopoulos-2017}. Here $\ln(R) / N = 0.052$ for $4 \times 4$ and $0.054$ for $6 \times 6$, so $R$ grows by $3.0 \times$, from 2.3 to 6.9; a direct asymptotic variance estimate gives a $3.3 \times$ sample increase, consistent with the bootstrap. As argued in the SM, at half filling we hypothesize that the number of samples grows to leading order as $\exp(c N)$ with $c < 0.1$, and we expect slower growth in dilute systems. The rate $c$ depends on $\theta$, whose optimum tracks $J$, but at fixed $\theta$ it appears insensitive to the reference MPS. After calibration, reaching the same relative precision requires about $30 \times$ more samples for $6 \times 6$ than for $4 \times 4$, reflecting calibration's reduced effectiveness. We attribute this in part to the fact that $\braket{\phi | (D - \braket{D})^2 | \phi} \propto N$.

\subsection{Demonstration on quantum hardware}
%
\begin{figure*}
  \centering
  \begin{minipage}[t]{0.325\textwidth}
    \makebox[\linewidth][l]{\textbf{(a)}}\par
    \includegraphics[width=\linewidth]{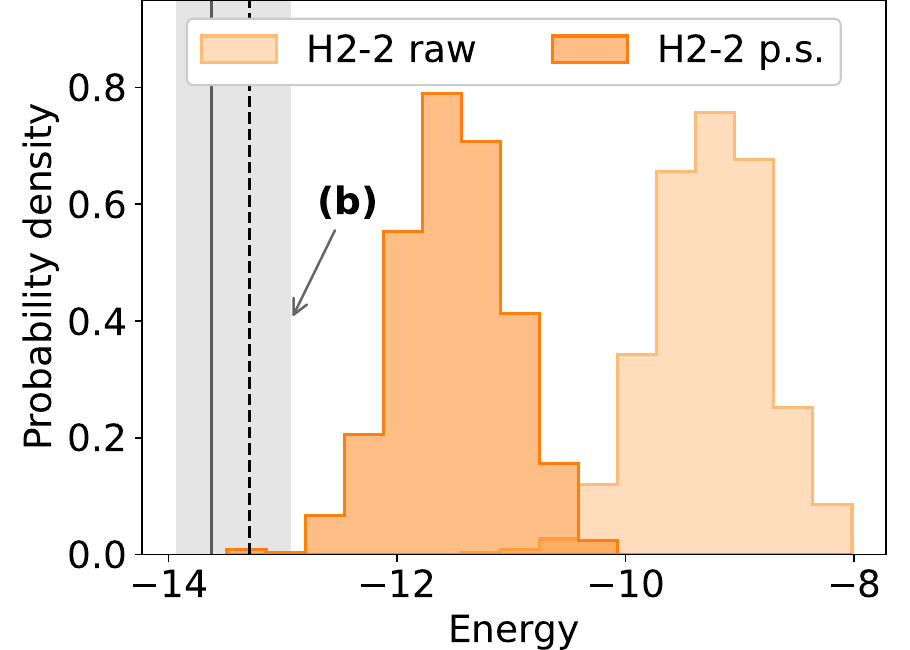}
  \end{minipage}
  \hfill
  \begin{minipage}[t]{0.325\textwidth}
    \makebox[\linewidth][l]{\textbf{(b)}}\par
    \includegraphics[width=\linewidth]{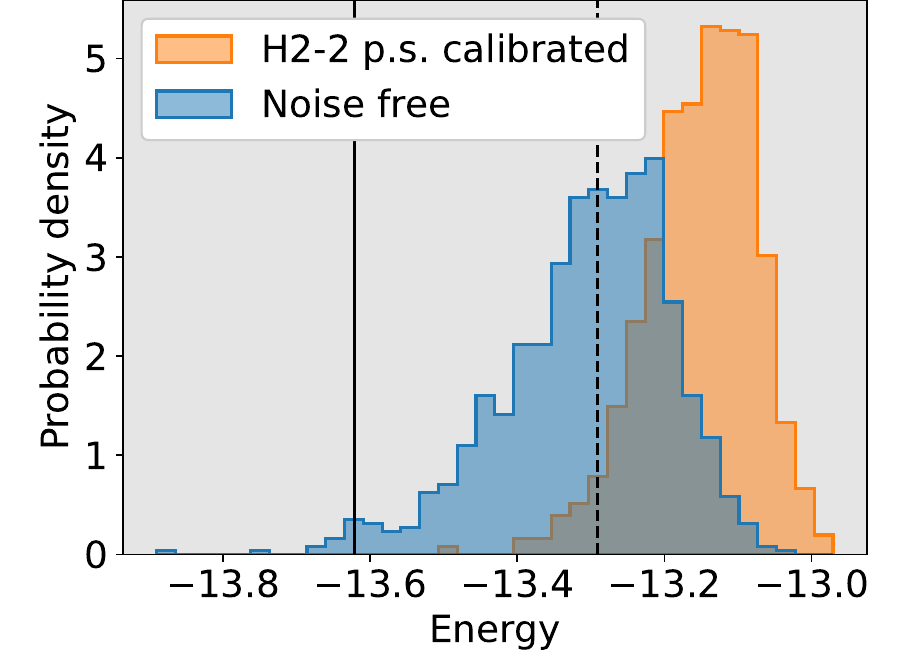}
  \end{minipage}
  \hfill
  \begin{minipage}[t]{0.325\textwidth}
    \makebox[\linewidth][l]{\textbf{(c)}}\par
    \includegraphics[width=\linewidth]{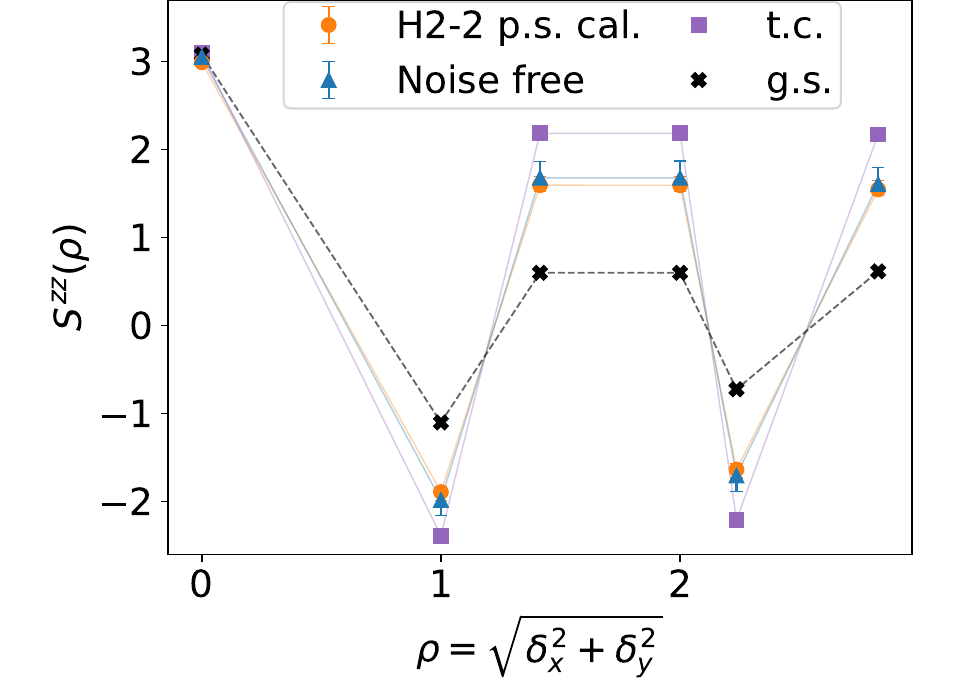}
  \end{minipage}
  \caption{\label{fig:H2-results}
    \textbf{(a)} Energy bootstrap distributions from raw and particle-number-postselected (p.s.) H2-2 data without calibration.
    \textbf{(b)} Postselected, calibrated hardware data and 1,000 calibrated noise-free experiments matched to the retained sample count in every measurement setting.
    \textbf{(c)} The $z$-component of the spin correlations from postselected and calibrated H2-2 data, matched-sample noise-free experiments, the biorthogonal transcorrelated MPS (t.c.), and the exact ground state (g.s.), averaged over displacements of length $\rho$.
    Solid and dashed lines in (a) and (b) mark the exact ground-state and full-sample noise-free energies.
  }
\end{figure*}

We implemented the bond-dimension-4 SB MPS on 32 qubits of the 56-qubit Quantinuum H2-2 quantum computer \cite{Quantinuum-H2-2-run}. Using the $\text{Iter}[D_i O_\text{all}]$ method of Ref.~\cite{Rudolph-Chen-2023}, we optimized a two-layer staircase preparation circuit with infidelity $1 - \left| \braket{\phi|U|0^{\otimes32}} \right|^2 = 1.4\times10^{-5}$ before compilation and excluding hardware error. After compilation into the native gate set, the complete diagonal-measurement circuit contains 555 two-qubit gates and 1,095 single-qubit gates. Each hopping-measurement circuit adds between one and three compiled two-qubit gates. The preparation and rotation circuits act only on nearest-neighbor qubits in a linear chain, and the hopping basis change can be folded into the rotation at no asymptotic cost, so the algorithm needs no two-dimensional connectivity.

We collected 12,250 diagonal and 23,616 hopping samples. The target MPS has fixed $N_\uparrow$ and $N_\downarrow$, which the basis rotation and measurements preserve. Discarding outcomes outside this sector therefore provides symmetry verification \cite{Bonet-Monroig-Sagastizabal-2018}. Postselection onto the target sector $N_\uparrow = N_\downarrow = 8$ retained 7,526 diagonal and 14,509 hopping samples. Figure~\ref{fig:H2-results}(a) compares the Gutzwiller-optimized energies from the raw and postselected samples, $-9.21 \pm 0.49$ and $-11.49 \pm 0.49$. Postselection improves the energy substantially, but both remain well above the full-sample noise-free benchmark of $-13.2914 \pm 0.0028$.

After calibrating the postselected distributions, we obtain $-13.14 \pm 0.07$, as shown in Fig.~\ref{fig:H2-results}(b). Calibration reduces the bootstrap standard error by $6.6 \times$ and the deviation from the full-sample noise-free energy by $12.2 \times$. This decrease is expected: the target real-space state has high overlap with the low-energy subspace, so typical stochastic hardware errors transfer weight to higher-energy states, while calibration limits their impact to higher-order Taylor terms. To separate residual hardware error from finite sampling, Fig.~\ref{fig:H2-results}(b) also shows 1,000 calibrated noise-free experiments matched to each hardware setting's retained sample count. The hardware estimate exceeds 95.2\% of these experiments, suggesting a residual hardware-dependent upward shift after postselection and calibration.

Figure~\ref{fig:H2-results}(c) tests observables beyond the optimized energy. We consider
\begin{equation*}
  S^{zz}(\rho) = \operatorname*{avg}_{|\V{\delta}| = \rho} \sum_{\V{x}} S^z_{\V{x}} S^z_{\V{x} + \V{\delta}} \ , \qquad S^z_{\V{x}} = \frac{n_{\V{x} \uparrow} - n_{\V{x} \downarrow}}{2} \ ,
\end{equation*}
which we evaluate from the diagonal samples. For every $\rho > 0$, the postselected, calibrated hardware estimates and the calibrated matched-sample noise-free estimates are closer to the exact correlations than are the biorthogonal transcorrelated results. For this bond-dimension-4 SB MPS, this ordering is opposite to that of the energies, Fig.~\ref{fig:classical-results}(a), showing that apparent energy accuracy, particularly for a nonvariational estimate, need not imply accurate observables. Since $4 S^{zz}(0) = N_\uparrow + N_\downarrow - 2 D$, $\rho = 0$ primarily measures the interaction energy, explaining why all three approximate methods are most accurate there. For these observables the discrepancy from the exact correlations is dominated by the reference state and correlator ansatz rather than hardware noise.

\section{Discussion}

Classically, combining a momentum-space MPS with a real-space diagonal correlator requires restricting one of them: VMC needs efficiently computable real-space reference amplitudes, and the transcorrelated method restricts the correlator and forgoes a variational energy. Our algorithm needs no reference amplitudes: the quantum processor prepares the momentum-space MPS, rotates it to real space, and samples it in bases fixed by the Hamiltonian alone. The correlator is therefore purely classical after measurement and can be changed or re-optimized without further quantum computation; any efficiently computable real-space diagonal form, including neural-network correlators \cite{Carleo-Troyer-2017,Gu-Li-Lin-2025}, is compatible. With exact expectation values, Hermiticity of $C$ guarantees a variational upper bound, and observables follow from the single state $C\ket{\phi}$ rather than a biorthogonal pair.

Among feedback-free variational hybrid algorithms (see SM\smcite), existing methods fix their trial spaces at measurement time and solve linear eigenvalue problems \cite{Kanno-Kohda-2026,McClean-Kimchi-Schwartz-2017,Parrish-McMahon-2019,Stair-Huang-2020,Klymko-Mejuto-Zaera-2022}. By contrast, our algorithm nonlinearly optimizes an arbitrary real-space diagonal correlator from the same measurement record, a nonconvex problem, though only a local minimum near the transcorrelated correlator is needed. Quantum-selected configuration interaction (QSCI) is an obvious comparison because it also repeatedly samples a prepared state \cite{Kanno-Kohda-2026}. Its projected energy retains a variational upper bound under both finite sampling and state-preparation errors, whereas our sampled ratio estimator has no strict finite-sample bound and calibration forfeits the guarantee under state-preparation error. Conversely, even when QSCI samples noiselessly from the exact ground state, it requires approximately $2 \times 10^8$ samples to reach the upper edge of the hardware 95\% bootstrap interval, nearly four orders of magnitude more than the $3.6 \times 10^4$ hardware samples collected here. Quantum Krylov subspace diagonalization (QKSD) instead pays in two-qubit gates. Even starting from a $\chi = 64$ real-space MPS and ignoring noise, controlled-time-evolution overhead, and state preparation, QKSD needs 544 two-qubit gates to match our hardware result, versus 464 for our SB rotation circuit, and 22{,}304 to match our $\chi = 16$ SI accuracy, 32 times the 696 of the corresponding rotation circuit.

Our experiment joins Fermi--Hubbard demonstrations on superconducting processors \cite{Stanisic-Bosse-2022} and trapped ions probing ground-state, dynamical, and pairing properties \cite{Nigmatullin-Hemerey-2025,Alam-Bosse-2025,Granet-Lin-2025}, but remains a proof of principle; quantum advantage would require larger lattices away from half filling, where auxiliary-field QMC has a sign problem \cite{Loh-Gubernatis-1990,Troyer-Wiese-2005}. Transcorrelated DMRG is more accurate for dilute systems, suggesting that small-bond-dimension MPSs may remain effective reference states \cite{Corbett-Miyake-2025}. Reaching this regime likely requires alternative MPS preparation \cite{Malz-Styliaris-Wei-2024}; momentum-space sparsity makes Ref.~\cite{Rupprecht-Wolk-2026} appealing. Tailoring the reference to the correlator may increase the benefit of richer correlators, as Fig.~\ref{fig:classical-results}(b) suggests. For the half-filled Gutzwiller correlator, our results support the hypothesis that the sample cost grows as $\exp(c N)$ with $c < 0.1$. How this cost changes with filling and richer correlators, and whether improved calibration or optimization techniques can lower it, remain open.


\begin{acknowledgments}
  Support is acknowledged from the U.S. Department of Energy, Office of Science, National Quantum Information Science Research Centers, Quantum Systems Accelerator (Award No. DE-SCL0000121).
  Support is acknowledged from the New Mexico State Legislature through House Bill 2 (2025), which provided us with fellowship support through the Quantum New Mexico Institute.
  We thank the UNM Center for Advanced Research Computing, supported in part by the National Science Foundation, for high-performance computing resources.
  OpenAI ChatGPT (GPT-5.6 Sol) and Anthropic Claude (Opus 5) were used to assist with scientific discussion, literature synthesis, code development, and manuscript preparation. All AI-assisted analysis and generated content were independently reviewed and verified by the authors.
\end{acknowledgments}

\bibliography{./references.bib}

\newpage 

\appendix

\section{End Matter}

\subsection{Transcorrelated DMRG}

The reference states are produced by transcorrelated DMRG \cite{Baiardi-Reiher-2020,Corbett-Miyake-2025}, which replaces the eigenvalue problem for $\ket{\psi} = C_\text{tc} \ket{\phi}$ with the non-Hermitian problem $H_\text{tc} \ket{\phi} = E \ket{\phi}$ for the reference state alone, where $H_\text{tc} = C_\text{tc}^{-1} H C_\text{tc}$. For the Fermi--Hubbard Hamiltonian the correlator is the Gutzwiller factor $C_\text{tc} = e^{J D}$, which for $J < 0$ suppresses the interaction energy, and the reference $\ket{\phi}$ is a momentum-space MPS. The nonvariational energy is then $E_\text{tc} = \braket{\phi | H_\text{tc} | \phi}$. We fix $J$ by the residual method \cite{Boys-Handy-1969c,Corbett-Miyake-2025}, choosing $J^*$ such that
\begin{equation*}
  r(J) = \braket{H_\text{tc}(J) D} - \braket{H_\text{tc}(J)} \braket{D}
\end{equation*}
vanishes, where expectation values are with respect to $\ket{\phi(J)}$. Viewing the explicitly correlated energy as a function of $J$, we find that $\ket{\phi(J^*)}$ is close to the optimal reference state, and that the optimized parameter $\theta^*$ is nearly equal to $J$ at every $J$ considered. The momentum-space form of $H_\text{tc}(J)$ and the biorthogonal construction required for observables are given in the SM\smcite.

\subsection{Spin-orbital ordering and quantum circuit}

The two orderings differ in how the momentum modes are laid out. The spin-interleaved (SI) ordering places the two spin modes at each momentum next to one another and arranges momentum pairs $(\V{k}, \V{k} + (\pi, \pi))$ by increasing $|\epsilon(\V{k})|$, where $\epsilon(\V{k})$ is the single-particle dispersion; it is the qubit version of the bipartite ordering of Ref.~\cite{Corbett-Miyake-2025}. The spin-blocked (SB) ordering places the two spin sectors on separate halves of the qubit register and reverses the order of the pairs in one sector, keeping the modes near the Fermi surface contiguous. The SI ordering localizes the strong entanglement between spin sectors and therefore gives faster DMRG convergence with increasing bond dimension. With the SB ordering, however, the Fourier transformation factorizes into two $N$-qubit unitaries. After compilation, the rotation circuit costs 464 and 696 two-qubit gates for the SB and SI orderings, respectively. The SI cost is less than twice the SB cost because many SI gates are bare fermionic swaps, which compile into a single two-qubit gate plus an implicit qubit relabeling \cite{Alam-Bosse-2025}.

The MPS preparation circuit uses the method of Ref.~\cite{Rudolph-Chen-2023}, and the basis rotation the matchgate compilation of Ref.~\cite{Zhao-Miyake-2024}; both are detailed in the SM\smcite. The linear connectivity requirement is a property of this staircase preparation, not of MPS preparation generally, since other constructions use ancilla control.

\subsection{Evaluation of correlated hopping term}

For any diagonal operator $A$, define $A(\cdot \,|\, b_k)$ as $A$ evaluated with $n_k$ replaced by $b_k \in \{ 0, 1 \}$, so that $A(h \,|\, b_k)$ is its value on configuration $h$, and extend this notation to multiple indices. Since $n_k^2 = n_k$,
\begin{equation*}
  a_k R = R(\cdot \,|\, 1_k) a_k \ , \qquad a_j^\dagger R = R(\cdot \,|\, 0_j) a_j^\dagger \ ,
\end{equation*}
with the corresponding left-acting identities
\begin{equation*}
  L a_j^\dagger = a_j^\dagger L(\cdot \,|\, 1_j) \ ,\qquad L a_k = a_k L(\cdot \,|\, 0_k) \ .
\end{equation*}
Therefore,
\begin{equation} \label{eq:L adag a R}
  L a^\dagger_j a_k R = a^\dagger_j a_k L(\cdot \,|\, 1_j, 0_k) R(\cdot \,|\, 0_j, 1_k) \ .
\end{equation}
For the explicitly correlated method, $L = R = C = C^\dagger$, so
\begin{equation*}
  C (a^\dagger_j a_k + a^\dagger_k a_j) C = (a^\dagger_j a_k + a^\dagger_k a_j) C(\cdot \,|\, 1_j, 0_k) C(\cdot \,|\, 0_j, 1_k) \ .
\end{equation*}
With the Jordan--Wigner mapping this turns into
\begin{equation*}
  \frac{X_j X_k + Y_j Y_k}{2} \left( \prod_{\ell = j + 1}^{k - 1} Z_\ell \right) C(\cdot \,|\, 1_j, 0_k) C(\cdot \,|\, 0_j, 1_k) \ .
\end{equation*}
In practice this means that we measure $(X X + Y Y) / 2$ on qubits $j$ and $k$ and $Z$ on every other qubit. From these $Z$ measurements we can reconstruct the Jordan--Wigner string outcome and use $n_\ell = (1 - Z_\ell) / 2$ for $\ell \notin \{ j, k \}$ to compute the correlator value.

\subsection{Calibration}

To incorporate a set of known expectation values $\{ \braket{O_r} \}_r$, we formulate a linearly constrained quadratic program that finds the minimum-norm multiplicative adjustment of the empirical distribution whose reweighting reproduces them exactly. Consider calibrating the diagonal samples. Each sample, $d_\ell$, is associated with a set of eigenvalues $\{ \lambda_{r \ell} \}$, and among all $M^{(d)}$ samples we find $X$ unique sets of eigenvalues each occurring with empirical frequency $f_x$. We reweight these frequencies into probabilities $p_x = \alpha_x f_x$ such that the quadratic program is
\begin{equation} \label{eq:quadratic-program}
  \begin{split}
    \V{\alpha}^* = \underset{{\V{\alpha}}}{\operatorname{arg \, min}} \, &\sum_{x = 1}^X (\alpha_x - 1)^2 \\
    \text{s.t.} \ &\sum_{x = 1}^X \lambda_{r x} \alpha_x f_x = \braket{O_r} \ .
  \end{split}
\end{equation}
The normalization constraint $\sum_x p_x = 1$ can be enforced by including the identity among $\{ O_r \}$. We solve this equality-constrained quadratic program, exploit its structure to reduce it to a linear system in the number of constraints, and then reject the solution if any resulting probability is negative. By seeking to minimize $(\alpha_x - 1)^2$, we reduce the chance of encountering negative probabilities because driving $p_x$ to zero costs unity in this objective, whereas the more natural objective $(p_x - f_x)^2$ charges only $f_x^2$ and so resists negative probabilities far more weakly. After solving for $\V{\alpha}^*$, if sample $d_\ell$ belongs to the $x$-th set, we assign it a probability $p^{(d)}_\ell = \alpha_x / M^{(d)}$.

The hopping terms are calibrated similarly, but the optimization is performed on four separate groups. Hopping terms are split by their spin and direction. Within each group, the translation symmetry of the momentum-space MPS (which has definite total momentum) is invoked to canonicalize the different hopping terms, which are then aggregated.

\begin{figure}
  \includegraphics[width=\columnwidth]{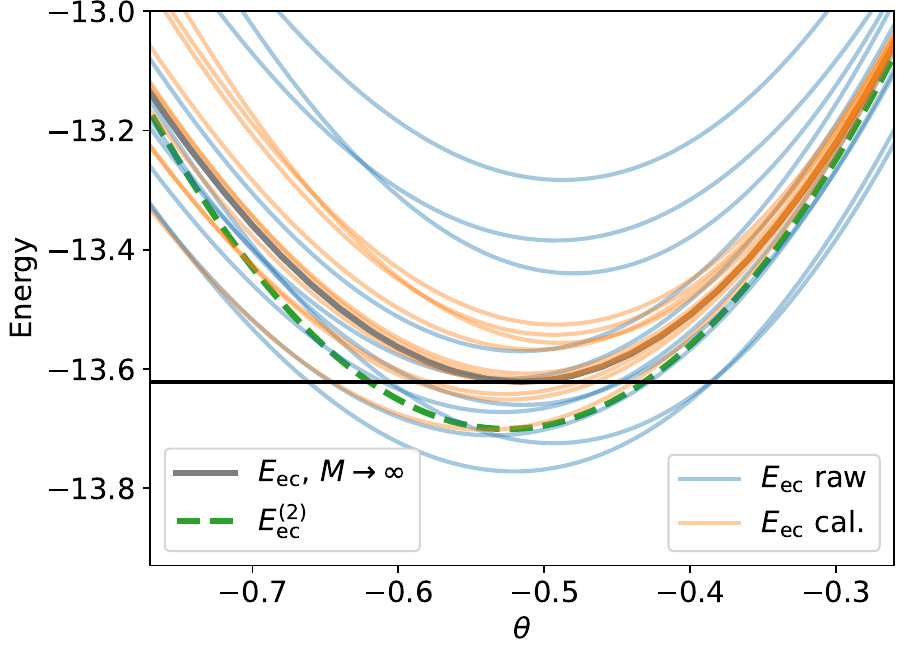}
  \caption{\label{fig:energy-vs-j}
    The explicitly correlated energy landscape with the Gutzwiller correlator, $e^{\theta D}$, for the $\chi = 1024$ SI MPS. Shown are the infinite-sample limit $E_\text{ec}, \, M \rightarrow \infty$, the second-order Taylor approximation $E^{(2)}_\text{ec}$, and ten randomly selected experiments each using 314,000 samples with and without calibration. The horizontal line is the exact ground-state energy.
  }
\end{figure}
Figure~\ref{fig:energy-vs-j} shows that calibration pins the finite-sample estimates to the exact curve near $\theta = 0$ and keeps them substantially more accurate through the variational minimum, $\theta^* \approx J = -0.515$, despite growing deviations as unconstrained higher-order terms become relevant. For the $J = 0$ non-transcorrelated MPSs, the small $\theta^*$ produces near-perfect collapse in this region, explaining the invisible error bars in Fig.~\ref{fig:classical-results}(b). Calibration and $E^{(2)}_\text{ec}$ share the lower-order moments, but the full quadratic approximation additionally uses $\braket{D D_r D_s}$ ($\braket{D^3}$ for Gutzwiller), omitted from calibration. $E^{(2)}_\text{ec}$ treats the moments as the entire energy model, whereas calibration imposes them as constraints on the sampled distributions, which then evaluate $E_\text{ec}(\V{\theta})$ without truncation. The two therefore agree near $\theta = 0$ and separate away from it, where $E^{(2)}_\text{ec}$ inherits the truncation error, overshooting the variational minimum and falling below $E_\text{gs}$.

\subsection{Sampling parameters}

Explicitly correlated energies in Figs.~\ref{fig:classical-results}(a) and \ref{fig:classical-results}(b) are computed from 31.4 million samples (12.2 million diagonal and 300{,}000 for each of the 64 hopping settings), except the $\chi = 512$ and $\chi = 1024$ MPSs, which use 157 million samples. The experiments in Fig.~\ref{fig:classical-results}(c) partition the same 157 million samples, ranging from 500,000 experiments of 314 samples to 50 experiments of 3.14 million samples. For the 314-sample experiments, hopping samples are calibrated in a single pool rather than per measurement setting, each of which would otherwise hold too few samples to constrain the calibration.

\subsection{Error bars}

Unless specified otherwise, error bars are 95\% percentile bootstrap confidence intervals from 1,000 stratified nonparametric replicates \cite{Efron-Tibshirani-1993}. Outcomes are resampled with replacement within each measurement setting, preserving its sample count; calibration and optimization are repeated. Figure~\ref{fig:classical-results}(c) shows central 68\% intervals across independent experiments; the noise-free points in Fig.~\ref{fig:H2-results}(c) show medians and central 95\% intervals across 1,000 independent matched-sample experiments. Text values report the data estimate $\pm$ the bootstrap standard error.

\subsection{Failed experiments}

Failures are concentrated in the smallest experiments of Fig.~\ref{fig:classical-results}(c): of the 500,000 experiments of 314 samples, 128,314 uncalibrated and 117,341 calibrated experiments failed. Every other configuration failed in at most 6\% of trials. The SM defines the failure modes and gives the complete breakdown\smcite.

\subsection{Software}

All MPS calculations use ITensors \cite{ITensors}, and we optimize the explicitly correlated energy using Newton's method with a trust region, as implemented in Optim \cite{Optim-2018}, with derivatives generated by Enzyme \cite{Enzyme-2020}. KrylovKit \cite{KrylovKit} supplies the eigensolvers and exponentiation used by the DMRG, configuration interaction, and time-evolution calculations.

\newpage

\section{Supplemental Material}

\subsection{Momentum-space transcorrelated Hamiltonian}

The transcorrelated eigenvalue problem is
\begin{equation*}
  H C \ket{\phi} = E C \ket{\phi} \quad \rightarrow \quad C^{-1} H C \ket{\phi} = E \ket{\phi} \ .
\end{equation*}
For the Gutzwiller correlator $C = e^{J D}$, transcorrelated DMRG requires a representation of the momentum-space transcorrelated Hamiltonian $H_\text{tc} = C^{-1} H C$ \cite{Luo-Alavi-2018,Baiardi-Reiher-2020}
\begin{align*}
  & H_\text{tc}(J) = H - \sum_{\V{p}, \V{q}, \V{k}, \sigma} \omega(\V{p}, \V{k}) a^\dagger_{\V{p} - \V{k}, \sigma} a^\dagger_{\V{q} + \V{k}, \bar{\sigma}} a_{\V{q}, \bar{\sigma}} a_{\V{p}, \sigma} \\
  &+ \sum_{\substack{\V{k}, \V{k}' \\ \V{s}, \V{q} \\ \V{p}, \sigma}} \gamma(\V{p}, \V{k}, \V{k}') a^\dagger_{\V{p} - \V{k}, \sigma} a^\dagger_{\V{q} + \V{k}', \bar{\sigma}} a^\dagger_{\V{s} + \V{k} - \V{k}', \bar{\sigma}} a_{\V{s}, \bar{\sigma}} a_{\V{q}, \bar{\sigma}} a_{\V{p}, \sigma} ,
\end{align*}
where
\begin{align*}
  \epsilon(\V{k}) &= 2 t (\cos k_x + \cos k_y) \ , \\
  \omega(\V{p}, \V{k}) &= \frac{1}{N} \left[ (e^J - 1) \epsilon(\V{p} - \V{k}) + (e^{-J} - 1) \epsilon(\V{p}) \right] \ , \\
  \gamma(\V{p}, \V{k}, \V{k}') &= 2 \frac{\cosh(J) - 1}{N^2} \epsilon(\V{p} - \V{k} + \V{k}') \ .
\end{align*}
Estimating observables beyond the energy requires the left eigenvector of the transcorrelated Hamiltonian, $\bra{\phi'} H_\text{tc} = E \bra{\phi'}$. With this in hand
\begin{equation*}
  \braket{\psi | O | \psi} = \frac{\braket{\phi' | C^{-1} O C | \phi}}{\braket{\phi' | \phi}} \ .
\end{equation*}
Since the left eigenvector is often harder to approximate than the right \cite{Luo-Alavi-2018,Baiardi-Reiher-2020}, these transcorrelated observable calculations are often less accurate than the energy itself.

\subsection{Transcorrelated Hamiltonian for a general correlator}

Consider the transcorrelated Hamiltonian $C^{-1} H C$ for a general two-electron diagonal correlator
\begin{equation*}
  C_{\V{\theta}} = \exp \left( \sum_{p < q} \theta_{p q} n_p n_q \right) \ .
\end{equation*}
Because $C$ is diagonal, $C^{-1} D C = D$, and $C^{-1} T C$ follows from Eq.~\eqref{eq:L adag a R}
\begin{equation} \label{eq:general transcorrelated method}
  C_{\V{\theta}}^{-1} a^\dagger_j a_k C_{\V{\theta}} = a^\dagger_j a_k \exp \left[ \sum_{p \notin \{j, k\}} (\theta_{p k} - \theta_{p j}) n_p \right] \ .
\end{equation}
Since $e^{\alpha n_p} = 1 + (e^\alpha - 1) n_p$, each nonzero term in the exponent of Eq.~\eqref{eq:general transcorrelated method} increases the many-body rank by one. This is why the Gutzwiller transcorrelated Hamiltonian, in which $\theta_{p k}$ only couples a spin orbital with the same orbital but opposite spin, is a three-electron operator. Consider a three-parameter correlator that, in addition to the on-site Gutzwiller term, includes nearest-neighbor same- and opposite-spin couplings. In two dimensions, there are 14 different terms, resulting in the transcorrelated Hamiltonian being a 15-electron operator. On its own this is not necessarily a problem; after all, the explicitly correlated Hamiltonian is an all-electron operator even with the Gutzwiller correlator. But for the Fermi--Hubbard Hamiltonian, the transcorrelated method is only beneficial in momentum space, and after rotating our $m$-electron transcorrelated real-space Hamiltonian into momentum space we have $O(N^{2 m - 1})$ terms (the $-1$ resulting from conservation of momentum). Even for the Gutzwiller transcorrelated momentum-space Hamiltonian, dealing with the $O(N^5)$ terms is the bottleneck in DMRG calculations.

One possibility is that this more complicated transcorrelated Hamiltonian could be approximated by discarding the higher many-body rank terms. Such truncations and related approximations are commonly employed in transcorrelated electronic-structure calculations \cite{Luo-2010,Luo-2011,Motta-Gujarati-Rice-2020,Kumar-Asthana-2022,Baiardi-Lesiuk-Reiher-2022}. Although the transcorrelated energy might be of little use, if the resulting state is better suited to the more general correlator, it could make a better reference for an explicitly correlated calculation.

\subsection{Details of quantum circuit construction}

We construct the MPS preparation circuit with the $\text{Iter}[D_i O_\text{all}]$ method of Ref.~\cite{Rudolph-Chen-2023}. A bond-dimension-2 approximation to the target MPS is converted into a staircase of two-qubit unitaries using the method of Ref.~\cite{Ran-2020}, then every gate is optimized by tensor-network contraction. For a gate $U_j$ with environment $E_j$ obtained by contracting the rest of the network, the fidelity between the state prepared by the circuit and the MPS is $\mathcal{F}(U_j) = |\tr(U_j E_j)|^2$. If $E_j = P \Sigma V^\dagger$, the update $U_j = V P^\dagger$ maximizes this objective while preserving unitarity. This process is repeated to add layers.

A particle-number-preserving Gaussian unitary has the form
\begin{align*}
  U(h) &= \exp \left( -i \sum_{pq} h_{p q} a_p^\dagger a_q \right) \ , \\
  U^\dagger(h) a_r U(h) &= \sum_s(e^{-ih})_{rs} a_s \ .
\end{align*}
It therefore implements any number-preserving single-particle basis rotation. Any Gaussian unitary can be compiled into a matchgate circuit of depth $O(N)$ with $O(N^2)$ gates. We use the compilation scheme of Ref.~\cite{Zhao-Miyake-2024}.

There are two benefits to measuring $a^\dagger_j a_k + a^\dagger_k a_j$ together instead of the individual Pauli terms. The first is a reduction in the number of measurement settings from two to one. The second is that the joint measurement preserves the number of electrons and allows us to perform postselection by discarding samples in which noise has changed the number of electrons. The downside is the addition of an extra nonlocal matchgate $U_\text{hop}$ required to diagonalize $(X X + Y Y) / 2$
\begin{equation*}
  U_\text{hop} = \frac{1}{2}
  \begin{pmatrix}
    2 & 0 & 0 & 0 \\
    0 & \sqrt{2} & -\sqrt{2} & 0 \\
    0 & \sqrt{2} & \sqrt{2} & 0 \\
    0 & 0 & 0 & 2
  \end{pmatrix} \ .
\end{equation*}
In principle the nonlocality can always be eliminated at no asymptotic cost by absorbing this particle-number-preserving matchgate into the definition of the fermionic rotation from momentum space. However, we do not employ this optimization here, since H2-2 has all-to-all connectivity.

\subsection{Classical MPS rotation}

\begin{figure}
  \includegraphics[width=\columnwidth]{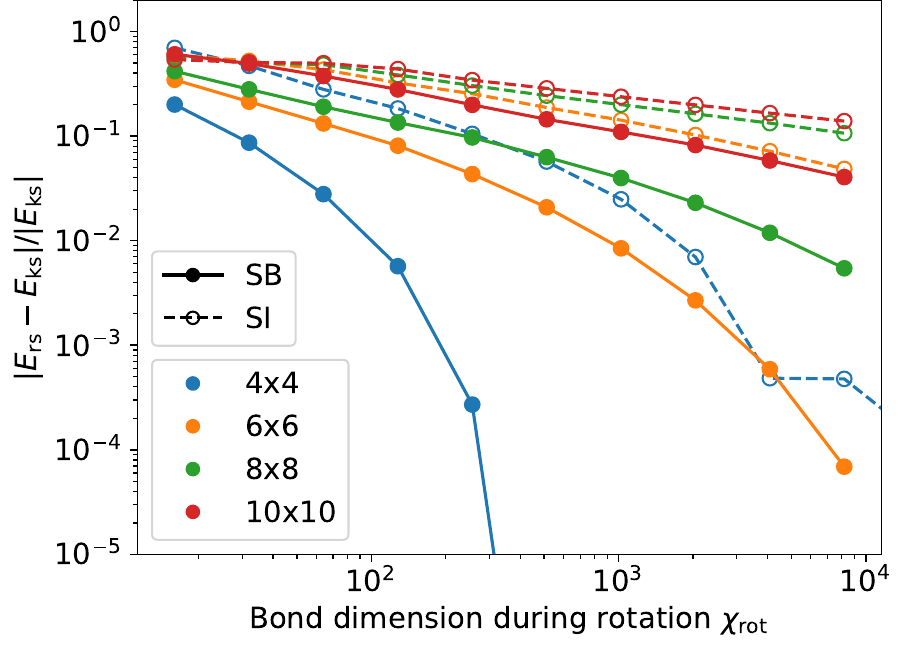}
  \caption{\label{fig:rotation-scaling}
    Relative energy error from applying the Fourier rotation classically to a $\chi = 16$ momentum-space MPS, truncating to bond dimension $\chi_\text{rot}$ after every matchgate, for half-filled lattices in the spin-blocked (SB) and spin-interleaved (SI) orderings. $E_\text{ks}$ is the energy of the momentum-space MPS and $E_\text{rs}$ that of the truncated real-space MPS.
  }
\end{figure}

Our algorithm uses the quantum processor to rotate the momentum-space MPS to real space. Here we assess whether this rotation could instead be performed classically. We apply the matchgate circuit described above directly to $\chi = 16$ MPSs in the SI and SB orderings, truncating to a maximum bond dimension $\chi_\text{rot}$ after each gate. Since the exact rotation preserves the energy, the deviation of the real-space energy $E_\text{rs}$ from the momentum-space energy $E_\text{ks}$ measures the truncation error.

The two orderings bound the required bond dimension differently. With the SB ordering, the rotation factorizes into two disjoint $N$-qubit rotations. Each effectively acts on $N + \log_2 \chi$ qubits, its $N$ physical qubits plus the $\chi$-dimensional bond to the other spin sector, so an exact rotation can be accomplished with $\chi_\text{rot} \le \sqrt{\chi 2^N}$. With the SI ordering, the circuit acts across the full $2N$-qubit chain and may require the maximal bond dimension $2^N$.

Figure~\ref{fig:rotation-scaling} reflects these bounds. The SB error decays faster than a power law over the accessible bond dimensions for the $4 \times 4$ and $6 \times 6$ lattices and begins to do so for $8 \times 8$. The $4 \times 4$ SI error enters the same regime once $\chi_\text{rot}$ is large relative to the system size, but for $6 \times 6$ and larger lattices this crossover is not reached and the SI error decays as a power law. The kink in the $4 \times 4$ SI curve near $\chi_\text{rot} = 2^{13}$ occurs because $E_\text{rs}$ crosses below $E_\text{ks}$; the discrepancy vanishes at larger $\chi_\text{rot}$.

At $\chi_\text{rot} = 2^{13}$, the SB error grows by nearly three orders of magnitude from $6 \times 6$ to $10 \times 10$. The SB and SI $10 \times 10$ calculations at this bond dimension each required more than two days of wall time on a 64-core Intel Xeon node, yet even the more accurate SB result retained a 4\% error. The quantum circuit instead performs the rotation exactly with $O(N^2)$ matchgates in depth $O(N)$.

\subsection{Infinite-sample calculations}

For the SI MPSs, where in real space the spin degrees of freedom corresponding to each site are assigned to adjacent qubits, the Gutzwiller correlator, $e^{\theta D}$, is a product of on-site factors $\operatorname{diag}(1, 1, 1, e^\theta)$ in the four-dimensional Hilbert space spanned by $\{ \ket{\emptyset}, \ket{\uparrow}, \ket{\downarrow}, \ket{\uparrow \downarrow} \}$, and therefore has an MPO representation of bond dimension 1. So we can exactly apply it to the real-space MPS to compute $E_\text{ec}(\theta) = \braket{\phi | e^{\theta D} H e^{\theta D} | \phi} / \braket{\phi | e^{2 \theta D} | \phi}$. This is what we refer to as the infinite-sample limit. We optimize this energy to within a convergence tolerance of $\Delta \theta = 0.001$. For the SB MPSs, $n_{\V{x} \uparrow} n_{\V{x} \downarrow}$ is not a local operator, and this computation is not feasible.

\begin{figure}
  \includegraphics[width=\columnwidth]{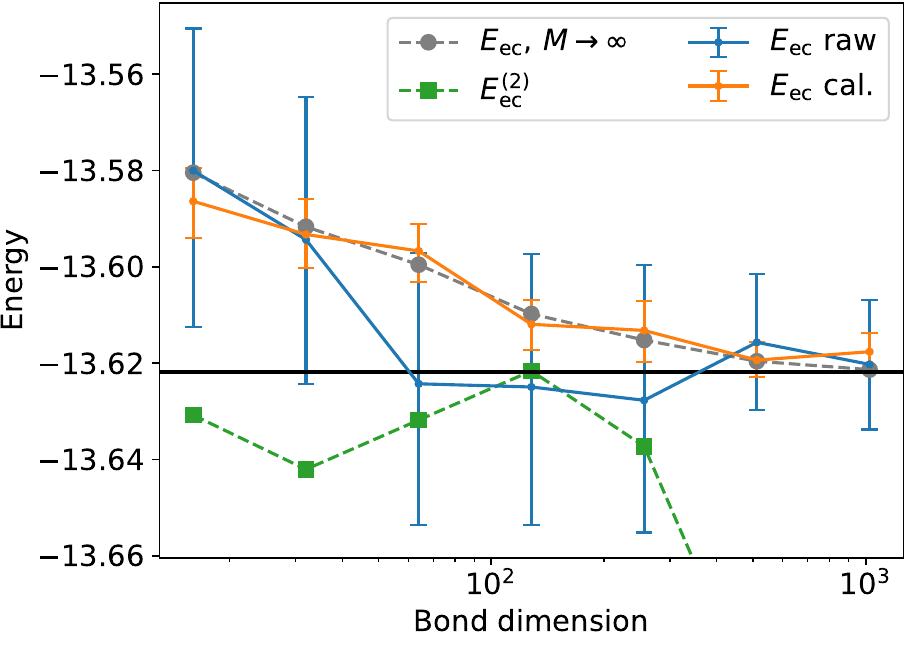}
  \caption{\label{fig:exact-gutzwiller}
    The infinite-sample explicitly correlated energy with the Gutzwiller correlator $E_\text{ec}, \, M \rightarrow \infty$, the second-order Taylor approximation, $E^{(2)}_\text{ec}$, and the finite-sample estimates with and without calibration for the SI MPSs. The horizontal line is the exact ground-state energy.
  }
\end{figure}
Figure~\ref{fig:exact-gutzwiller} is a zoomed-in version of Fig.~\ref{fig:classical-results}(a), with the addition of the infinite-sample limit and the uncalibrated explicitly correlated energies. The calibrated energies track the exact curve far more closely and carry significantly smaller uncertainty than the raw data. Although the second-order Taylor approximation nearly reproduces the ground-state energy at $\chi = 128$, the agreement is coincidental: its accuracy should be judged against the infinite-sample limit. The drastic decrease in accuracy of the second-order approximation at $\chi = 512$ and 1024 is due to the decrease of $J^*$ to $-0.51$ and $-0.515$, respectively, compared with a minimum of $-0.485$ for the smaller bond dimensions shown here. Moreover, we see that the infinite-sample energy converges to the ground-state energy smoothly as the bond dimension is increased, demonstrating that transcorrelated DMRG approximates the transcorrelated ground state well.

\begin{table*}[t]
  \caption{\label{tab:failed-experiments}
    Failed experiments and bootstrap replicates, broken down by failure mode. ``Optimization'' counts runs where the explicitly correlated energy diverged or otherwise failed to converge, ``$\theta$ bounds'' counts runs whose optimum fell outside $\theta_r \in [-3, 3]$, and ``Calibration'' counts runs where the linear system implicit in Eq.~\eqref{eq:quadratic-program} admitted no solution or gave negative probabilities. Trials are bootstrap replicates for Fig.~\ref{fig:classical-results}(b) and independent experiments for Fig.~\ref{fig:classical-results}(c).
  }
  \begin{ruledtabular}
    \begin{tabular}{llrrrrr}
      Figure & Configuration & Trials & Optimization & $\theta$ bounds & Calibration & Total \\
      \hline
      \ref{fig:classical-results}(b) & 59 parameters, $\chi = 4$    & 1,000   & 7       & 0   & 0   & 7       \\
                                     & 59 parameters, $\chi = 32$   & 1,000   & 34      & 0   & 0   & 34      \\
                                     & 103 parameters, $\chi = 4$   & 1,000   & 1       & 0   & 0   & 1       \\
                                     & 103 parameters, $\chi = 8$   & 1,000   & 56      & 0   & 0   & 56      \\
      \hline
      \ref{fig:classical-results}(c) & 314 samples, uncalibrated    & 500,000 & 127,874 & 440 & 0   & 128,314 \\
                                     & 314 samples, calibrated      & 500,000 & 116,594 & 93  & 654 & 117,341 \\
                                     & 3,140 samples, uncalibrated  & 50,000  & 13      & 0   & 0   & 13      \\
                                     & 3,140 samples, calibrated    & 50,000  & 5       & 0   & 0   & 5       \\
    \end{tabular}
  \end{ruledtabular}
\end{table*}
%
\subsection{Failed experiments} \label{sec:failed-experiments}

Table~\ref{tab:failed-experiments} tallies the failed experiments and bootstrap replicates by failure mode. We classify as optimization failures runs in which the explicitly correlated energy diverges or the optimizer otherwise fails to converge. We separately reject runs whose optimum lies outside the parameter bounds $\theta_r \in [-3,3]$, and calibrated runs for which the constraints in Eq.~\eqref{eq:quadratic-program} admit no solution or produce a negative probability.

Aside from the pathological 314-sample experiments, all the failures are due to non-convergent (almost exclusively divergent) optimization of the explicitly correlated energy. These failures arise because finite-sample estimates of the numerator and denominator of Eq.~\eqref{eq:explicitly-correlated-energy} are obtained from different measurement distributions and therefore need not retain the consistency of an exact Rayleigh quotient. The mechanism is particularly transparent for the Gutzwiller correlator $C=e^{\theta D}$. Let $D_{\min}^{(d)}$ be the smallest number of doublons occurring in the finite diagonal sample. As $\theta\rightarrow-\infty$, the estimated normalization then scales as
\begin{equation*}
  \braket{C^2} \sim e^{2\theta D_{\min}^{(d)}} \ .
\end{equation*}
A hopping outcome with negative eigenvalue and reconstructed doublon numbers $D_1$ and $D_2$ in the two correlation factors contributes to the numerator as $-e^{\theta(D_1+D_2)}$, up to a positive prefactor. If
\begin{equation*}
  D_1 + D_2 < 2 D_{\min}^{(d)} \ ,
\end{equation*}
this negative contribution decays more slowly than the sampled normalization, so the estimated energy diverges to $-\infty$. For example, if the diagonal sample contains no doublon-free configurations, $D_{\min}^{(d)}=1$, even a negative hopping contribution with one doublon across its two correlator factors gives an asymptotic ratio proportional to $-e^{-\theta}$. This pathology is purely a finite-sampling effect and cannot occur for the exact expectation values, for which Eq.~\eqref{eq:explicitly-correlated-energy} is a Rayleigh quotient bounded below by the ground-state energy. However, the more flexible a correlator is, the more it can exploit these finite-sample discrepancies between the diagonal and hopping distributions and escape to $-\infty$, consistent with the higher divergence rate observed for the 59- and 103-parameter correlators.

Not every consequence of this inconsistency is a failure. Because the finite-sample objective is not a Rayleigh quotient, an individual bootstrap replicate can also fall slightly below the exact ground-state energy without diverging. Among the successful replicates of Fig.~\ref{fig:classical-results}(b), these excursions are concentrated in the most accurate configurations, where the residual variational error is smallest relative to the finite-sampling spread; there they reach 45\% of replicates, and across all configurations no replicate falls more than 0.06\% below $E_\text{gs}$.

\subsection{Optimization procedure}

For a fixed set of diagonal and hopping samples, Eqs.~\eqref{eq:diagonal-estimators} and \eqref{eq:hopping-estimator} define a deterministic nonlinear objective $E_\text{ec}(\V{\theta})$. We optimize this objective using Newton's method with a trust region, as implemented in Optim \cite{Optim-2018}, with derivatives generated by Enzyme \cite{Enzyme-2020}. In preliminary tests, first-order gradient-based methods were substantially more likely to find the divergent trajectories described in the preceding subsection, whereas the curvature information and restricted steps of the trust-region Newton method made the optimization more robust.

Each optimization is initialized from the transcorrelated correlator used to generate the reference MPS. For the Gutzwiller ansatz this sets $\theta=J$; for the more general correlators, the Gutzwiller coefficient is initialized to $J$ and all additional coefficients to zero. Bootstrap replicates are optimized independently from this same transcorrelated starting point rather than initialized from the optimum obtained with the full sample set. Thus each replicate repeats the complete optimization procedure and probes the sensitivity of the optimized correlator itself to finite sampling.

This nonlinear optimization distinguishes the present method from QSCI and QKSD, whose classical post-processing reduces to an eigenvalue problem solvable in polynomial time. Here the objective is generally nonconvex, and nonconvex global minimization is NP-hard \cite{Murty-Kabadi-1987}. Nevertheless, nonlinear optimization of this kind is routine in physics, most notably in VMC, and our method does not require the global minimum. Indeed, for the finite-sample objective the global minimum need not even be physically meaningful: a local minimum in the basin of the transcorrelated correlator can give an accurate approximation to the infinite-sample optimum even when a distant region of parameter space contains a spurious direction along which $E_\text{ec} \rightarrow -\infty$.

Optimization over a fixed finite sample set also gives the problem a natural statistical-learning interpretation. The sampled energy $E_\text{ec}(\V{\theta})$ plays the role of an empirical objective, while the infinite-sample energy is the corresponding population objective. Increasingly expressive correlators can exploit sample-specific discrepancies that do not generalize to the population distributions, with the divergent directions described above providing an extreme example. This suggests that standard tools for controlling generalization, such as held-out validation data, cross-validation, or regularization, could be adapted to improve robustness. The independent-sample validation of the 103-parameter correlator described in the following section provides one such generalization test.

\subsection{Correlator features}

The 59-parameter and 103-parameter correlators have the form of Eq.~\eqref{eq:explicitly-correlated-energy}, with the former tailored to the SB MPSs and the latter to the SI MPSs. Their features were selected by an LLM-assisted heuristic search using the available samples, without reference to the exact ground-state energy. Features were added and pruned to lower the energy consistently across several MPS bond dimensions while minimizing variation among independent nonoverlapping subsets of the samples and maintaining low optimization-failure rates.

To assess the generalization of the 103-parameter feature set, we collected four additional independent sample sets for the benchmark $\chi = 16$ SI MPS, with 31.4 million samples each, as in the original calculation. The features were held fixed and only their parameters were reoptimized on each sample set. The 103-parameter correlator lowers the energy relative to the one-parameter Gutzwiller correlator by 0.0229 on the feature-selection data and by 0.0204 on average across the four validation calculations. The difference, 0.0026, has a 95\% bootstrap interval of $[-0.0015, 0.0068]$ and is therefore not statistically resolved. Thus the improvement from the selected features largely generalizes to independent samples.

The absolute energy of the original sample realization is nevertheless particularly favorable: as shown in Fig.~\ref{fig:exact-gutzwiller}, its Gutzwiller estimate lies 0.0059 below the infinite-sample value, whereas all four validation estimates lie above it. Correspondingly, the upper 95\% relative-error bounds of the four 103-parameter validation calculations range from 0.20\% to 0.25\%, compared with 0.139\% for the original calculation. We therefore conservatively use the largest validation bound, 0.25\%, as the $\chi = 16$ SI target in the QSCI and Krylov comparisons below. The complete definition of the two correlators follows.

We first define a shared set $\mathcal{B}$ of 33 translation-summed local features. Nine are density pairs $P_{\Delta,s} = \frac{1}{4} \sum_{\V{x}, \sigma} \sum_{\V{\delta} \in \Delta, \eta = \pm 1} n_{\V{x} \sigma} n_{\V{x} + \eta \V{\delta}, \sigma \oplus s}$, where $s = 0$ and 1 denote equal and opposite spins, respectively, and
\begin{align*}
  (\Delta, s) =& (\{ (0, 0) \}, 1), \\
  & (\{ (1, 0), (0, 1) \}, s), (\{ (1, 1), (1, -1) \}, s), \\
  & (\{ (2, 0), (0, 2) \}, s), (\{ (2, 1), (1, 2) \}, s), \quad s = 0, 1 \ .
\end{align*}
The remaining 24 features count patterns on every $2 \times 2$ plaquette or $2 \times 3$/$3 \times 2$ rectangle. Number plaquette sites by $k = x + 2 y$, with $x, y \in \{ 0, 1 \}$. For singly occupied plaquettes, form the four-bit spin word $b = \sum_{k = 0}^3 2^k n_{k \downarrow}$ and retain the five channels $\min(b, 15 - b) \in \{ 0, 2, 3, 4, 7 \}$. For charge words, use the digits $q_k = n_{k \uparrow} + n_{k \downarrow} \in \{ 0, 1, 2 \}$ and the base-three word $\sum_{k = 0}^3 3^k q_k$; the six channels have canonical representatives $\{ 5, 8, 13, 14, 16, 24 \}$ under the eight square symmetries and $q_k \mapsto 2 - q_k$. On a $2 \times 3$ rectangle, again number sites by $k = x + 2 y$. Singly occupied six-bit spin words give nine channels $\{ 3, 5, 6, 7, 13, 18, 21, 22, 25 \}$ after minimization under the two rectangle reflections and global spin reversal. Four further channels count a single charge defect, $q_k \ne 1$, at an end or in the middle of the long direction, or one holon and one doublon at nearest-neighbor or non-nearest-neighbor separation. The plaquette and rectangle counts are divided by 4 and 8, respectively.

The 59-parameter correlator contains all 33 features of $\mathcal{B}$ together with 26 Schmidt-cut moments tied to the SB qubit order. Let $X_\ell = \sum_{j = 1}^{\ell} n_j - \ell / 2$ be the particle-number fluctuation across the cut after qubit $\ell$, where $n_j$ is the occupation of qubit $j$ in the SB ordering. The additional features are the quadratic moments $X_\ell^2 / 4^2$ for $\ell = 2, \dots, 14$ and $\ell = 18, \dots, 30$. The remaining cuts are excluded because $X_\ell^2$ is constant within the target sector: $X_{16} = X_{32} = 0$ since $N_\uparrow = N_\downarrow = 8$, and for $\ell = 1, 15, 17, 31$ the cut differs from a fixed-number cut by a single qubit, so $X_\ell = \pm 1 / 2$.

The 103-parameter correlator uses 29 features from $\mathcal{B}$, omitting spin-plaquette channel 7, spin-rectangle channels 7 and 13, and the single middle charge-defect channel, together with 74 complete rectangle-motif counts. Encode a $2 \times 3$ rectangle by $p = \sum_{\delta_x = 0}^1 \sum_{\delta_y = 0}^2 4^{\delta_x + 2 \delta_y} (n_{\V{x} + \V{\delta}, \uparrow} + 2 n_{\V{x} + \V{\delta}, \downarrow})$. Map each $p$ to the minimum integer in its orbit under the two rectangle reflections and spin exchange, then sort and index the distinct nonzero representatives from one. Separate features count both rectangle orientations without further normalization for the following set $\mathcal{I}$, where $a{:}b$ denotes every integer from $a$ through $b$:
\begin{align*}
  \mathcal{I} = \{&31, 47{:}48, 51, 61, 96, 100, 115, 127, 172, \\[-2pt]
  &174, 181, 183, 222, 230, 238, 243, 252{:}253, 258{:}259, \\[-2pt]
  &261, 265, 268, 271{:}274, 284, 309, 326, 329, 344{:}345, \\[-2pt]
  &347, 351, 357, 368, 373, 398, 400, 410{:}411, 416{:}418, \\[-2pt]
  &425, 427{:}429, 431, 435, 437, 441, 445{:}446, 453, 464, \\[-2pt]
  &466, 470, 486, 489, 491{:}492, 517, 524, 526, 534, 543, \\[-2pt]
  &546, 550, 553, 560, 566 \} \ .
\end{align*}

\subsection{System size sampling scaling}

\begin{figure}
  \includegraphics[width=\columnwidth]{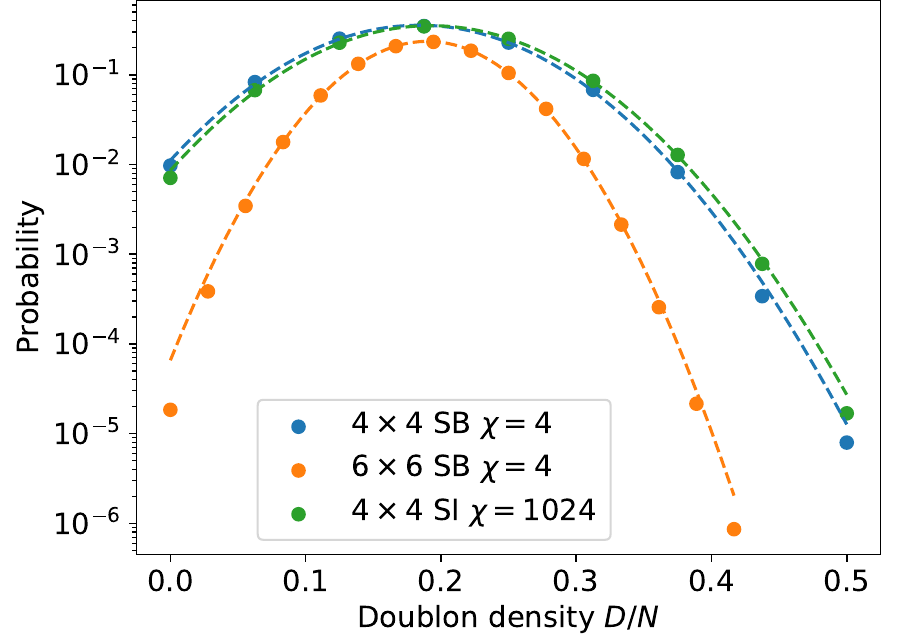}
  \caption{\label{fig:doublon-distribution}
    Empirical probability distribution of the doublon density $D / N$ at half filling for transcorrelated momentum-space MPSs of bond dimension $\chi$ and the spin-blocked (SB) and spin-interleaved (SI) qubit orderings for the $4 \times 4$ and $6 \times 6$ systems. The points are probability masses at integer doublon counts and sum to one. Dashed curves are continuous Gaussian approximations with unit integral in $D$, using $\braket{D}$ and $\braket{D^2}$ calculated directly from the momentum-space MPSs. The $4 \times 4$ MPSs were rotated to real space exactly, whereas the $6 \times 6$ MPS was truncated to a bond dimension $\chi_\text{rot} = 2^{13}$ at an energy error of 0.02\% relative to the initial MPS; its sampled mean and variance per site agree with the MPS moments in Table~\ref{tab:doublon-moments} to three decimal places.
  }
\end{figure}

The correlator can favor configurations that are rare in the reference state, increasing the sampling cost of $E_\text{ec}(\V{\theta})$ as this mismatch grows with system size. The cost depends on fluctuations of both the numerator and the sampled normalization in Eq.~\eqref{eq:explicitly-correlated-energy}, including their covariance. To assess this growth, we rotated a $\chi = 4$ SB transcorrelated momentum-space MPS ($J = -0.46$) for the $6 \times 6$ system to real space. In doing so we truncated the bond dimension to $2^{13}$ after applying every gate. Although the rotation is inexact, we obtained a 0.02\% error in the real-space energy relative to the initial momentum-space energy. We then took 24.4 million diagonal samples and 600{,}000 samples for each of the 144 hopping settings. After calibration, with the Gutzwiller correlator we obtained an energy error of $1.83 \pm 0.06\%$ relative to the AFQMC ground-state energy $E_\text{gs} = -30.865(9)$ \cite{Qin-Shi-Zhang-2016} at $\theta^* = -0.4397$. Compare this to the $4 \times 4$ system, where, with $3.5 \times$ fewer samples from the $\chi = 4$ SB MPS ($J = -0.455$), we obtain a $2.43 \pm 0.02\%$ relative energy error at $\theta^* = -0.4397$. Thus, despite the additional samples, the bootstrap standard error of the $6 \times 6$ relative energy error is $3.1 \times$ larger.

In order to explain this drastic loss in sample efficiency, in Fig.~\ref{fig:doublon-distribution} we plot the probability distribution of the doublon density $D / N$ in the reference state before applying the correlator for the two $\chi = 4$ SB MPSs mentioned above. For the $6 \times 6$ system, sampling a configuration with few doublons is much less likely than for the $4 \times 4$ system. It is precisely these low-doublon configurations that the Gutzwiller correlator amplifies, and the amplification also increases with the system size. At $\theta^*$, reweighting by $\exp(2 \theta^* D)$ increases the probability of $D = 0$ by $8.3 \times$ and $139 \times$ for $4 \times 4$ and $6 \times 6$, respectively, while the probability of $D / N < 0.1$ is amplified by $4.0 \times$ and $13.1 \times$. We also plot the doublon distribution of the $\chi = 1024$ SI MPS with $J = -0.515$ for the $4 \times 4$ system to show that this behavior is insensitive to the specific MPS used. The bulk of each distribution is well approximated by a Gaussian with the moments of $D$ obtained from the MPS (dashed curves); the largest deviation is in the far tails, e.g., the $6 \times 6$ probability of $D = 0$ is $3.6 \times$ below the Gaussian, precisely where reweighting is most sensitive.

\begin{figure}
  \includegraphics[width=\columnwidth]{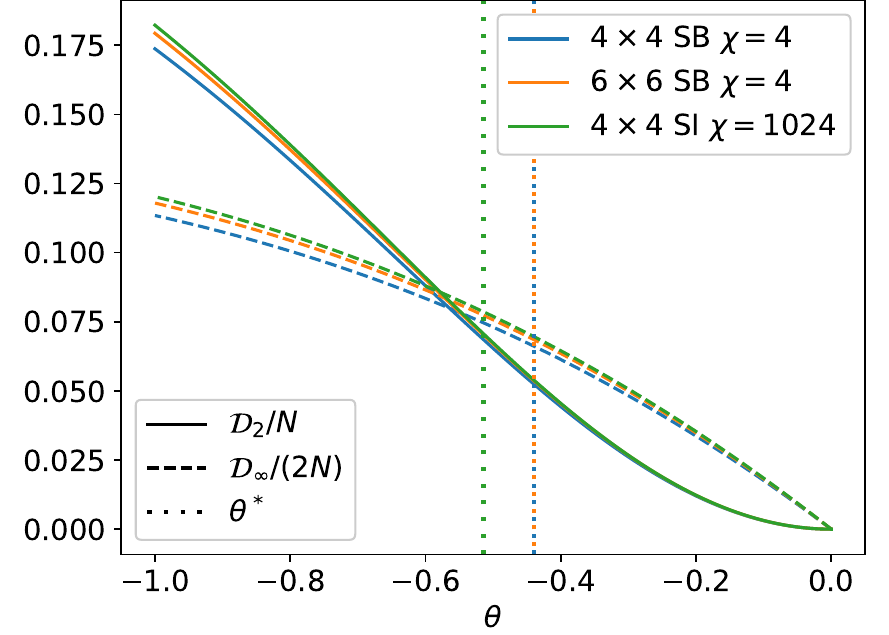}
  \caption{\label{fig:doublon-cumulant-scaling}
    Solid and dashed curves show the order-two and order-infinity R\'enyi divergences between the Gutzwiller correlated and reference doublon distributions, normalized as $\mathcal{D}_2 / N$ and $\mathcal{D}_\infty / (2 N)$, respectively. The references are transcorrelated momentum-space MPSs: the $\chi = 4$ spin-blocked (SB) MPSs for $4 \times 4$ and $6 \times 6$, and the $\chi = 1024$ spin-interleaved (SI) MPS for $4 \times 4$. Color-matched dotted vertical lines mark the corresponding optima. The full-sample calibrated bootstrap means and standard deviations are $\theta^* = -0.4397 \pm 0.0011$, $-0.4397 \pm 0.0038$, and $-0.51415 \pm 0.00065$, respectively.
  }
\end{figure}

The diagonal contribution is a self-normalized importance-sampling estimate \cite{Owen-2013,Agapiou-Papaspiliopoulos-2017}. Let $\Pr(d) = \left| \braket{d | U_\text{FT} | \phi} \right|^2$ be the reference real-space distribution. For any diagonal correlator $C$, define its normalization $Z_C = \braket{\phi | C^2 | \phi}$, the correlated distribution $\Pr_C(d) = C(d)^2 \Pr(d) / Z_C$, and expectation values $\braket{O}_C = \braket{\phi | C O C | \phi} / Z_C$. Here $C = C_{\V{\theta}}$; samples drawn from $\Pr(d)$ estimate these correlated expectation values, which for diagonal $O$ such as $D$ are averages over $\Pr_C(d)$,
\begin{align} \label{eq:importance-sampling-definition}
  \braket{D}_C &= \sum_d \Pr\nolimits_C(d) D(d) \nonumber \\
  &\approx \hat{D}_C = \frac{\sum_\ell p^{(d)}_\ell C^2(d_\ell) D(d_\ell)}{\sum_\ell p^{(d)}_\ell C^2(d_\ell)} \ .
\end{align}
To quantify the sampling cost of Eq.~\eqref{eq:importance-sampling-definition}, consider random variables $X$ and $Y$ with unbiased sample means $\hat{X}$ and $\hat{Y}$, from which we estimate $\braket{X} / \braket{Y}$ by $\hat{X} / \hat{Y}$. Linearizing this ratio about $(\braket{X}, \braket{Y})$ gives, to leading order in the inverse sample size,
\begin{equation*}
  \operatorname{Var} \left( \frac{\hat{X}}{\hat{Y}} \right) \simeq \frac{1}{\braket{Y}^2} \operatorname{Var} \left( \hat{X} - \frac{\braket{X}}{\braket{Y}} \hat{Y} \right) \ .
\end{equation*}
Taking $X = C^2 D$ and $Y = C^2$, the numerator and denominator of Eq.~\eqref{eq:importance-sampling-definition}, with $p^{(d)}_\ell = 1 / M^{(d)}$, yields
\begin{align} \label{eq:importance-sampling-full}
  \operatorname{Var}(\hat{D}_C) &\simeq \frac{\braket{\phi | C^4 (D - \braket{D}_C)^2 | \phi}}{M^{(d)} Z_C^2} \nonumber \\
  &= \frac{R}{M^{(d)}} \braket{(D - \braket{D}_C)^2}_{C^2} ,
\end{align}
where
\begin{align} \label{eq:importance-sampling-overhead}
  R &= \frac{Z_{C^2}}{Z_C^2} = \sum_d \frac{\Pr_C(d)^2}{\Pr(d)} \nonumber \\
    &= \exp \mathcal{D}_2 \left( \Pr\nolimits_C \,\Vert\, \Pr \right) ,
\end{align}
where $\mathcal{D}_2$ is the order-two R\'enyi divergence. Since the R\'enyi divergence measures how far the correlated distribution is from the reference, more samples will be required the further apart they are. In Fig.~\ref{fig:doublon-cumulant-scaling} we plot $\mathcal{D}_2 / N$ with the Gutzwiller correlator as a function of $\theta$ for the $\chi = 4$ SB MPSs of the $4 \times 4$ and $6 \times 6$ systems and the $\chi = 1024$ SI MPS of the $4 \times 4$ system. At their optima $\theta^* \approx -0.440$, $\mathcal{D}_2 \approx 0.05 N$ for the two $\chi = 4$ SB systems.

For the two $\chi = 4$ SB systems, treating the second factor in Eq.~\eqref{eq:importance-sampling-full} as extensive, and combining $R$ with an extensive number of measurement settings and ground-state energy, predicts a
\begin{equation} \label{eq:sampling-simple-ratio}
  \frac{M^{(6 \times 6)}}{M^{(4 \times 4)}} \simeq \frac{R^{(6 \times 6)}}{R^{(4 \times 4)}} \frac{\xi + 144}{\xi + 64} \frac{16}{36} = 2.34
\end{equation}
times increase in the number of samples to obtain a given relative precision, where $\xi = 122 / 3$ is the fixed ratio of diagonal samples to samples per hopping pair. Interpolating the raw bootstrap curves of Fig.~\ref{fig:ci-width-vs-samples} in logarithmic coordinates at fixed relative width gives a $3.35$--$3.44 \times$ increase where both budgets exceed $10^6$ samples, in qualitative agreement with this simple model. A direct asymptotic variance estimate, Eq.~\eqref{eq:sampling-direct-ratio}, which applies the same linearization to the full energy estimator, gives a $3.32 \times$ increase, consistent with the observed behavior; there, the sampled normalization contributes a term growing as $N^2 R$.

With calibration, the same interpolation gives about $28$--$32 \times$ over relative widths $2.5 \times 10^{-3}$--$1.2 \times 10^{-2}$, due to the decreased effectiveness of calibrating the $6 \times 6$ system. The diagonal calibration fixes $\braket{D}$ and $\braket{D^2}$, and hence the terms through second order in
\begin{equation*}
  e^{2 \theta D} = e^{2 \theta \braket{D}} \left[ 1 + 2 \theta \delta D + 2 \theta^2 \delta D^2 + O(|2 \theta \delta D|^3) \right] ,
\end{equation*}
where $\delta D = D - \braket{D}$ and the constant prefactor cancels in the energy ratio. With $\sigma^2(D) = \braket{\delta D^2}$, the uncontrolled higher-order terms are set by $2 |\theta| \sigma(D)$, which grows as $\sqrt{N}$ at fixed $\theta$; at $\theta^*$ it is $0.98$ for $4 \times 4$ and $1.48$ for $6 \times 6$.

\begin{table}
  \caption{\label{tab:doublon-moments}
    Expected number of doublons ($\braket{D}$), doublon variance $\sigma^2(D)$, and third central moment $\mu_3$, normalized by the number of sites $N$. All entries are calculated from momentum-space MPS moments, rather than sample moments. The first two rows are the $\chi = 4$ MPSs with the spin-blocked qubit ordering. The second and third sets of rows are the $\chi = 8$ MPSs from Ref.~\cite{Corbett-Miyake-2025} for the half-filled and dilute systems. These $\chi = 8$ MPSs use a local Hilbert space of dimension 4 and are therefore comparable to $\chi = 16$ qubit MPSs.
  }
  \scriptsize
  \setlength{\tabcolsep}{2pt}
  \begin{ruledtabular}
    \resizebox{\columnwidth}{!}{
    \begin{tabular}{rrrrr}
      System         & $N_e$ & $\braket{D} / N$ & $\sigma^2(D) / N$ & $10^3 \mu_3 / N$ \\ \hline
      $4 \times 4$   & 16    & 0.184            & 0.078             & 3.2  \\
      $6 \times 6$   & 36    & 0.190            & 0.079             & 2.5  \\
      \hline
      $6 \times 6$   & 36    & 0.192            & 0.078             & 2.5  \\
      $8 \times 8$   & 64    & 0.196            & 0.079             & 2.0  \\
      $10 \times 10$ & 100   & 0.202            & 0.079             & 0.8  \\
      $12 \times 12$ & 144   & 0.207            & 0.078             & $-0.1$ \\
      \hline
      $6 \times 6$   & 24    & 0.110            & 0.054             & 7.7  \\
      $8 \times 8$   & 26    & 0.041            & 0.028             & 11.8 \\
      $8 \times 8$   & 28    & 0.048            & 0.031             & 12.0 \\
      $8 \times 8$   & 44    & 0.118            & 0.056             & 7.4  \\
    \end{tabular}
    }
  \end{ruledtabular}
\end{table}

Although we cannot accurately rotate more complicated MPSs, we can compute the first three moments. In Table~\ref{tab:doublon-moments} we give $\braket{\phi | D | \phi}$, $\sigma^2(D)$, and $\mu_3 = \braket{\phi | (D - \braket{D})^3 | \phi}$ normalized by the lattice size $N$ for the two $\chi = 4$ SB MPSs used above and every $\chi = 8$ MPS from our previous work \cite{Corbett-Miyake-2025}. For the half-filled systems, although the doublon density has not yet converged, the variance is essentially unchanged at $\sigma^2(D) \approx 0.078 N$ while the normalized third central moment $\mu_3 / N$ tends toward zero. The agreement between the normalized R\'enyi divergence curves for the $4 \times 4$ and $6 \times 6$ systems, along with the normalized moments for the larger half-filled systems (aside from $\braket{D}$, which does not enter into $\mathcal{D}_2$), motivates the hypothesis that $\mathcal{D}_2 \approx 0.05 N$ at $\theta \approx -0.440$. The divergence curves for the $\chi = 1024$ SI MPS in Fig.~\ref{fig:doublon-cumulant-scaling} closely follow those from the $\chi = 4$ SB MPSs at fixed $\theta$, supporting robustness across these references. At its more negative optimum $\theta^* \approx -0.514$, the SI MPS has $\mathcal{D}_2 \approx 0.071 N$. We therefore hypothesize that the half-filled sampling cost scales as $\exp(c N) \operatorname{poly}(N)$ with $c < 0.1$; the polynomial factor includes the prefactor of $R$ in the estimator variance, which we examine with the direct variance estimate below. Predicting scaling away from half filling is more difficult, since we only have the moments. We tried approximating $\mathcal{D}_2$ using a third-order cumulant expansion and a discrete Gaussian based on the known spectrum of $D$ for the $6 \times 6$ system. Neither approximation was accurate enough to give us confidence in applying them to the dilute systems. At fixed system size and the same small negative $\theta$, the reduced $\sigma^2(D)$ and larger positive $\mu_3$ of the dilute systems lower the divergence through third order,
\begin{equation*}
  \mathcal{D}_2(\theta) = 4 \theta^2 \sigma^2(D) + 8 \theta^3 \mu_3 + O(\theta^4) \ .
\end{equation*}
This favors dilute systems in the weak-correlator regime, but does not establish their sampling cost at the optimized $\theta^*$, which can itself change with filling.

\subsubsection{Comparison with block encoding}

An alternative to reconstructing correlated observables in post-processing is to apply the correlator coherently with a block encoding \cite{Andras-Yuan-2019}. For the Gutzwiller correlator with $\theta \leq 0$, $\norm{C_\theta} \leq 1$, so there is a unitary $U_C$ with $U_C \ket{\phi} \ket{0} = C_\theta \ket{\phi} \ket{0} + \ket{\perp}$, where $(I \otimes \bra{0}) \ket{\perp} = 0$. Postselecting the ancilla on $\ket{0}$ therefore succeeds with probability $Z_C$, and the postselection overhead $Z_C^{-1}$, provided $\Pr(d)$ has support at $D = 0$, is the exponential of the order-infinity R\'enyi divergence between $\Pr_C(d)$ and $\Pr(d)$. By monotonicity of the R\'enyi divergence, $R = e^{\mathcal{D}_2} \leq e^{\mathcal{D}_\infty} = Z_C^{-1}$. For the systems in Fig.~\ref{fig:doublon-cumulant-scaling}, $(\mathcal{D}_\infty - \mathcal{D}_2) / N \approx 0.08$--$0.09$ at their respective $\theta^*$. If these rates persist, the postselection overhead grows exponentially faster than $R$. With amplitude amplification the block-encoding postselection overhead scales as $e^{\mathcal{D}_\infty / 2}$, and we plot $\mathcal{D}_\infty / (2 N)$ in Fig.~\ref{fig:doublon-cumulant-scaling}. Its crossings with $\mathcal{D}_2 / N$ depend weakly on the reference state and occur near $\theta = -0.57$; $\mathcal{D}_2 < \mathcal{D}_\infty / 2$ at all three optima. In our previous work, for all the half-filled systems the residual optimization yielded $J > -0.525$. If the optima remain on this side of the crossings at larger system sizes, even amplitude amplification does not close the exponential-overhead gap with our method. This is not to say that our method is superior to block-encoding the correlator. The constant factors involved in the sampling and postselection likely strongly favor block encoding. Additionally, when used as an input to a long phase estimation circuit, repeating the block encoding until success could be a small price for the increased ground-state overlap.

\subsubsection{Direct variance estimate}

\begin{table*}
  \caption{\label{tab:feedback-free-comparison}
    Resources beyond initial-state preparation required to reach the upper edge of the H2-2 95\% bootstrap interval (4.40\% relative error) and the $\chi = 16$ SI benchmark (0.25\%). Parameters and compiled two-qubit gates exclude the initial MPS and its preparation; the QKSD and UVQPE rows start from a $\chi = 64$ real-space SI MPS, whereas this work uses the $\chi = 4$ SB and $\chi = 16$ SI momentum-space MPSs. The parameter entries count correlator coefficients for this work, retained CI coefficients for QSCI, and the Krylov dimension $m$ for QKSD and UVQPE; thus, unlike Fig.~\ref{fig:ci}(a), they do not include MPS tensor entries. The zero QSCI gate count denotes computational-basis measurement with no two-qubit gates beyond state preparation. For QKSD and UVQPE, $N_\text{T}$ is the number of sequential second-order Trotter steps, each charged 272 two-qubit gates; these gate counts omit the coherent overhead of phase-sensitive measurement. The explicitly correlated and QSCI sample counts are literal samples, whereas $M$ is the favorable model-based sampling-resource proxy of Eqs.~\eqref{eq:krylov-base-cost} and \eqref{eq:krylov-sampling-costs} derived from Ref.~\cite{Lee-Lee-Huh-2024}, not a measured sample count. The ``min. $N_\text{T}$'' rows minimize the Trotter depth $N_\text{T}$ within $M \leq 10^{12}$, and the ``min. $N_\text{T} M$'' rows minimize the work proxy $N_\text{T} M$ over the scanned contour points; the QKSD ``min. $N_\text{T} M$ ($s = 1$)'' row minimizes $N_\text{T} M$ with no temporal subsampling.
  }
  \begin{ruledtabular}
    \begin{tabular}{llrrrrrr}
      Method & Operating point    & \multicolumn{3}{c}{H2-2: 4.40\%}                   & \multicolumn{3}{c}{$\chi=16$ SI: 0.25\%} \\ \cline{3-5} \cline{6-8}
             &                    & Parameters         & 2Q gates & Sampling resource  & Parameters         & 2Q gates & Sampling resource \\ \hline
      This work
             & ---                & 1                  & 464      & $3.59 \times 10^4$ & 103                & 696      & $3.14 \times 10^7$ \\
      Sampled QSCI \cite{Kanno-Kohda-2026}
             & ---                & $2.54 \times 10^7$ & 0        & $2.32 \times 10^8$ & $8.58 \times 10^7$ & 0        & $1.06 \times 10^{10}$ \\
      \hline
      QKSD \cite{Parrish-McMahon-2019,Stair-Huang-2020}
             & min. $N_\text{T}$  & 3                  & 544      & $1.61 \times 10^6$ & 83                 & 22{,}304  & $6.63 \times 10^{11}$ \\
             & min. $N_\text{T}M$ & 6                  & 4{,}080  & $1.10 \times 10^2$ & 60                 & 112{,}336 & $5.02 \times 10^3$ \\
             & min. $N_\text{T}M$ ($s=1$) & 12         & 2{,}992  & $2.25 \times 10^2$ & 226                & 61{,}200  & $5.17 \times 10^4$ \\
      \hline
      UVQPE \cite{Klymko-Mejuto-Zaera-2022}
             & min. $N_\text{T}$  & 3                  & 816      & $4.08 \times 10^7$ & 237                & 64{,}464  & $8.75 \times 10^{11}$ \\
             & min. $N_\text{T}M$ & 25                 & 6{,}800  & $6.19 \times 10^1$ & 449                & 122{,}128 & $1.10 \times 10^5$ \\
    \end{tabular}
  \end{ruledtabular}
\end{table*}

The simple model of Eq.~\eqref{eq:sampling-simple-ratio} assumes that, apart from the factor $R$, the estimator variance is extensive. To test this assumption, we evaluate the leading-order (delta-method) variance of the uncalibrated estimate $\hat{E}$ of $E = E_\text{ec}(\theta)$ at fixed $\theta$, with every measurement setting sampled independently. The diagonal samples estimate both the diagonal part of the numerator of Eq.~\eqref{eq:explicitly-correlated-energy} and the shared normalization $Z_C$, so each contributes the influence $C(d)^2 [U D(d) - E] / Z_C$. Each hopping setting $s$ estimates only a numerator term, with influence $W_s(h) T_s(h) / Z_C$, where $W_s(h) = C(h \,|\, 1_j, 0_k) C(h \,|\, 0_j, 1_k)$ is the correlator factor in Eq.~\eqref{eq:hopping-estimator}. Consequently,
\begin{align*}
  \operatorname{Var}(\hat E) &\simeq
  \frac{V^{(d)}}{M^{(d)}} + \frac{V^{(s)}}{M^{(s)}}, \\
  V^{(d)} &= \frac{\operatorname{Var}[C^2 (U D - E)]}{Z_C^2}, \\
  V^{(s)} &= \frac{1}{Z_C^2} \sum_s \operatorname{Var}(W_s T_s),
\end{align*}
where each single-shot variance is evaluated over the reference distribution for its measurement setting: $\Pr(d)$ for diagonal samples and $\Pr_s(h)$ for hopping samples. With $M^{(d)} = \xi M^{(s)}$ and $M = (\xi + 4 N) M^{(s)}$, this becomes $\operatorname{Var}(\hat E) \simeq V (\xi + 4 N) / M$ with $V = V^{(d)} / \xi + V^{(s)}$. The sampled distributions give $V^{(4 \times 4)} = 69.36$ and $V^{(6 \times 6)} = 670.62$ in units of $t^2$, so with the reference energies of Fig.~\ref{fig:ci-width-vs-samples} the total-sample ratio at fixed relative precision is
\begin{equation} \label{eq:sampling-direct-ratio}
  \frac{M^{(6 \times 6)}}{M^{(4 \times 4)}} =
  \frac{\xi + 144}{\xi + 64} \frac{V^{(6 \times 6)}}{V^{(4 \times 4)}}
  \left( \frac{E_\text{gs}^{(4 \times 4)}}{E_\text{gs}^{(6 \times 6)}} \right)^2 = 3.32 \ .
\end{equation}
Near the variational minimum, reoptimizing $\theta$ changes the energy only at second order, so the fixed-$\theta$ variance captures the leading bootstrap fluctuations. Because $V$ is evaluated from the same distributions that the bootstrap resamples, this is a consistency check on the bootstrap rather than an independent prediction.

The simple model predicts each variance coefficient to grow by $(R^{(6 \times 6)} / R^{(4 \times 4)}) \times 36 / 16 = 6.7$ from $4 \times 4$ to $6 \times 6$. The hopping coefficient $V^{(s)}$ grows by $7.2 \times$, close to this, whereas the diagonal coefficient $V^{(d)} / \xi$ grows by $30 \times$. Expanding $V^{(d)}$ isolates the contribution of the sampled normalization,
\begin{align*}
  V^{(d)} &= \frac{\operatorname{Var}(C^2 U D)}{Z_C^2} + E^2 (R - 1) \\
  &- \frac{2 E \operatorname{Cov}(C^2 U D, C^2)}{Z_C^2} ,
\end{align*}
where $\operatorname{Var}(C^2) / Z_C^2 = R - 1$ follows from Eq.~\eqref{eq:importance-sampling-overhead}. Since $E \propto N$, the second term scales as $N^2 R$ rather than the $N R$ of the simple model.

\subsection{Other feedback-free variational hybrid algorithms}

We call a ground-state algorithm \emph{feedback-free variational hybrid} if a classical computer performs nontrivial processing of the quantum measurements, which excludes phase estimation, but does not iteratively update the circuit run on the quantum processor, which excludes the variational quantum eigensolver (VQE). We further require that, in at least some regime, its energy estimates be variational, meaning upper bounds on the exact ground-state energy. Here we briefly outline four prominent examples---quantum-selected configuration interaction (QSCI), quantum subspace expansion (QSE), quantum Krylov subspace diagonalization (QKSD), and unitary variational quantum phase estimation (UVQPE)---before analyzing QSCI, QKSD, and UVQPE in detail on our benchmark system; QSE is set aside for reasons given below. Table~\ref{tab:feedback-free-comparison} compares the resources required by the three methods examined quantitatively with those of the present algorithm.

The first row of Table~\ref{tab:feedback-free-comparison} is the present algorithm. The compared methods fix their trial space at measurement time, so enlarging or replacing the ansatz requires new quantum data, whereas our correlator is chosen, enlarged, and nonlinearly optimized afterwards from the same samples; only observables outside the span of the measured bases require new measurements. At both targets, our method uses fewer literal samples than sampled QSCI and fewer compiled two-qubit gates than every reported Krylov method. QKSD and UVQPE, like our finite-sample ratio estimator, lose a strict bound under sampling error, and the Trotterized UVQPE calculations reported here are not variational even at infinite sampling because they diagonalize an approximate propagator; among the methods in the table, only QSCI retains the Ritz upper bound under both finite sampling and state-preparation error.

\subsubsection{Quantum-selected configuration interaction}

QSCI samples an approximate ground state in the computational basis and retains the observed configurations as a classical selected-CI space \cite{Kanno-Kohda-2026}. It then constructs the projected Hamiltonian classically and solves an eigenvalue problem (EVP) for its lowest eigenvalue, which is a Ritz upper bound; finite sampling and state-preparation errors change the selected configurations but do not destroy that guarantee.

\subsubsection{Quantum subspace expansion}

QSE forms the trial space $\{ O_j \ket{\phi} \}_{j = 0}^{m - 1}$ from a fixed set of excitation operators, measures $S_{j k} = \braket{\phi | O_j^\dagger O_k | \phi}$ and $H_{j k} = \braket{\phi | O_j^\dagger H O_k | \phi}$, and solves the generalized eigenvalue problem (GEVP) $H \V{c} = E S \V{c}$ \cite{McClean-Kimchi-Schwartz-2017,Takeshita-Rubin-2020}. The generally non-Hermitian products in these matrix elements are evaluated by decomposing them into measurable Hermitian real and imaginary components. Exact, mutually consistent matrices give a Ritz upper bound, and imperfect preparation only changes the trial subspace. Independent finite-sample errors can make $H$ and $S$ inconsistent, removing the guarantee.

We do not investigate QSE further because, unlike the other algorithms in this section, it is unlikely to offer a quantum advantage for a reference state with an efficient MPS representation. The common fermionic ansatz uses single excitations $a^\dagger_k a_\ell$, and applying an individual excitation string to an MPS of bond dimension $\chi$ does not increase that bond dimension. The same holds for higher-order excitation strings. QSE may still improve the energy; it simply does not need a quantum processor to do so.

\subsubsection{Quantum Krylov subspace diagonalization}

QKSD solves the same GEVP as QSE but with $O_j = \left(e^{-i \tau H} \right)^j$ for a real time step $\tau$ \cite{Parrish-McMahon-2019,Stair-Huang-2020}. Because $e^{-i \tau H}$ can grow the entanglement of the reference state, the trial state of QKSD need not remain efficiently representable as an MPS, and therefore escapes the argument that sets QSE aside. Unlike fermionic excitation operators, $e^{-i \tau H}$ cannot generally be expanded efficiently in a local operator basis, so phase-sensitive overlap protocols are used to estimate the projected matrix elements. A standard Hadamard-test implementation introduces an ancilla and controlled time evolution. Ancilla-free, fidelity-based alternatives replace controlled evolution with circuits involving the preparation of $\ket{\phi}$ and its inverse, together with phase-reference superpositions \cite{Cortes-Gray-2022}. For the Fermi--Hubbard Hamiltonian, the vacuum is an exact eigenstate in a different particle-number sector, making superpositions of the form $(\ket{\phi} + e^{i \theta} \ket{0}) / \sqrt{2}$ straightforward to prepare. With exact evolution, both $H$ and $S$ are Toeplitz matrices with $O(m)$ independent entries, where $m$ is the basis size. Under Trotterized evolution, $S$ remains Toeplitz, but $H$ generally does not, so all $O(m^2)$ Hamiltonian matrix elements must be estimated. Because the true Hamiltonian is still projected into the Trotter-generated trial space, the Trotterization need not accurately approximate the true evolution.

\subsubsection{Unitary variational quantum phase estimation}

UVQPE uses the same real-time Krylov basis but projects the one-step propagator $U = e^{-i \tau H}$ rather than $H$, solves the GEVP $U \V{c} = \lambda S \V{c}$, and extracts energies as $E = -\arg(\lambda) / \tau$ on an appropriate phase branch \cite{Cortes-Gray-2022,Klymko-Mejuto-Zaera-2022}. This preserves the Toeplitz structure of the projected $U$ matrix, even under Trotterization. However, because the procedure estimates the eigenphases of the Trotterized unitary, the Trotterization must accurately approximate the true evolution. The sufficient condition $\tau W < \pi$, where $W = E_\text{max} - E_\text{min}$ is the spectral width of $H$, also preserves the variational upper bound, whereas placing the spectrum within a single phase period requires only $\tau W < 2\pi$. The scan below imposes the latter condition.

\subsection{Quantum-selected configuration interaction}

\begin{figure*}[t]
  \centering
  \begin{minipage}[t]{0.49\textwidth}
    \makebox[\linewidth][l]{\textbf{(a)}}\par
    \includegraphics[width=\linewidth]{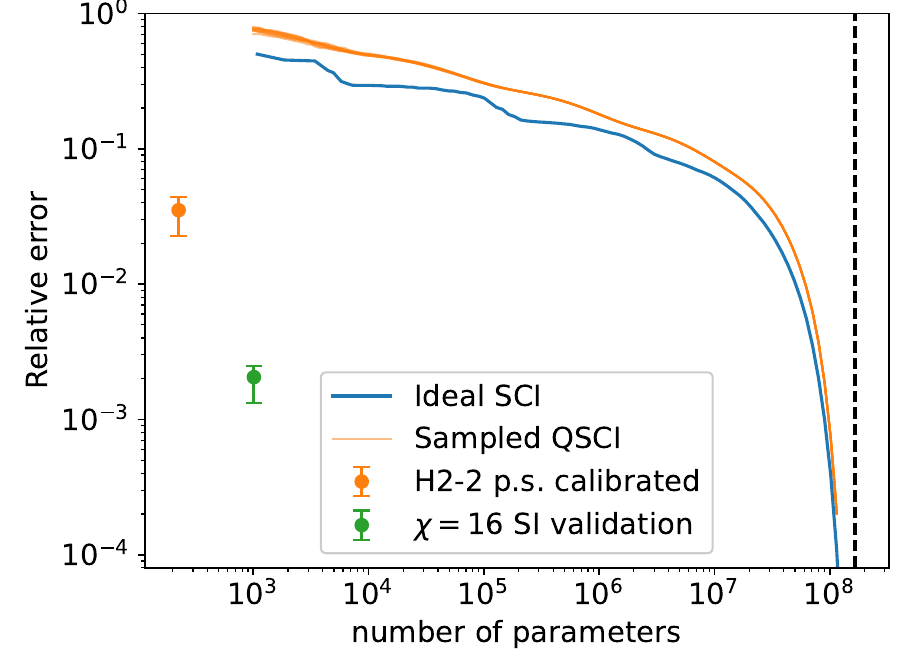}
  \end{minipage}
  \hfill
  \begin{minipage}[t]{0.49\textwidth}
    \makebox[\linewidth][l]{\textbf{(b)}}\par
    \includegraphics[width=\linewidth]{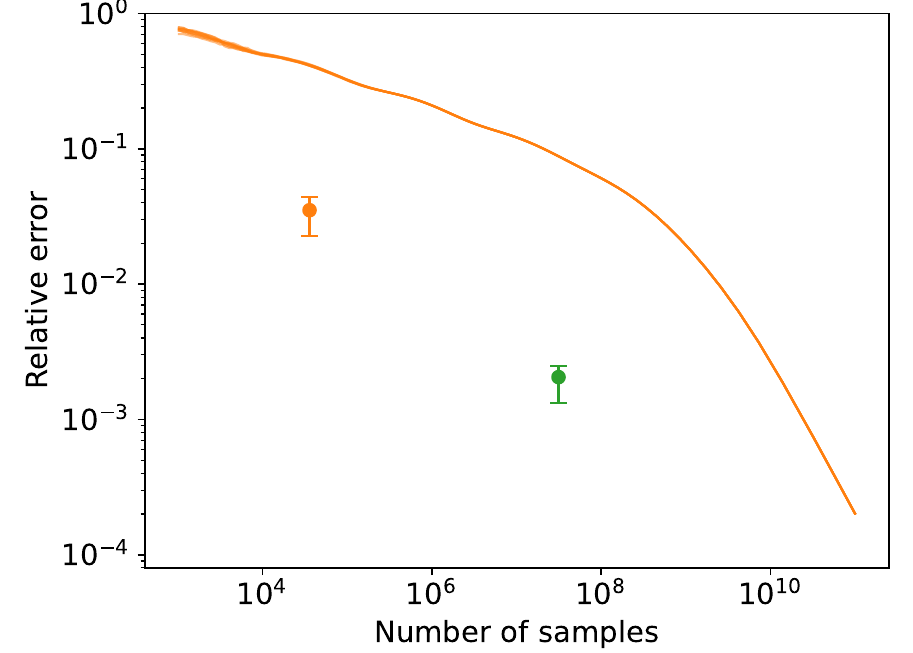}
  \end{minipage}
  \caption{\label{fig:ci}
    Relative energy error against the two resources that the explicitly correlated and quantum-selected configuration interaction ans\"atze both consume: the number of parameters, \textbf{(a)}, and the number of quantum samples, \textbf{(b)}. In both panels the points are the postselected, calibrated H2-2 explicitly correlated energy and the $\chi = 16$ SI energy with the 103-parameter correlator for the most conservative validation calculation, whose upper 95\% relative-error bound is 0.25\%.
    \textbf{(a)} For the configuration interaction (CI) curves the parameter count is the CI subspace dimension, one variational coefficient per retained configuration. The Ideal curve ranks configurations by decreasing exact ground-state probability; the Sampled curves are ten independent with-replacement trajectories from that distribution. For the explicitly correlated points the parameter count is that of the MPS--correlator ansatz, counting symmetry-allowed MPS tensor entries as parameters: 225 for the $\chi = 4$ SB reference measured on H2-2 (224 tensor entries and one Gutzwiller parameter) and 1012 for the $\chi = 16$ SI reference (909 tensor entries and 103 correlator parameters). The vertical dashed line marks the full fixed-particle-number sector.
    \textbf{(b)} The same sampled trajectories against the cumulative number of samples from the exact ground-state distribution. The probability-ranked curve has no associated sample count and is therefore not shown. The explicitly correlated points are placed at the number of samples used: the 35,866 samples from H2-2 and the $3.14 \times 10^7$ noise-free samples in the validation calculation.
  }
\end{figure*}

Though often paired with VQE, QSCI can also be run as a feedback-free variational hybrid algorithm with built-in error mitigation \cite{Kanno-Kohda-2026,Robledo-Moreno-Motta-2025,Sugisaki-Kanno-2025,Weaving-Mingare-2026}. Like our algorithm, it repeatedly samples a state prepared on a quantum processor, so the differences between the two are worth setting out in detail.

QSCI uses a quantum state as an importance sampler for a classical variational calculation \cite{Kanno-Kohda-2026}. Let $\mathcal{S}_B = \operatorname{span} \{\ket{x_1}, \ldots, \ket{x_B}\}$ be the subspace spanned by $B$ distinct configurations, and define the projected Hamiltonian matrix $(H_B)_{i j} = \braket{x_i | H | x_j}$. The QSCI energy is its lowest eigenvalue,
\begin{equation*}
  E_\text{QSCI}(B) = \lambda_\text{min}(H_B) \ .
\end{equation*}
Restricting $H$ to a subspace can only raise its lowest eigenvalue, so $E_\text{QSCI}$ is a variational upper bound even when the sampled state and measurements are imperfect. This robustness comes with two distinct costs: the selected subspace can itself be large, and sampling with replacement repeatedly returns configurations already in the subspace. Both limitations have been identified in molecular QSCI calculations \cite{Reinholdt-Ziems-2025}, while a recent study of Heisenberg and Hubbard models found large configuration spaces even when configurations were included in decreasing order of their exact ground-state probability \cite{Gaberle-Jattana-2026}.

We isolate these intrinsic costs using the exact ground state of the half-filled $4 \times 4$ Fermi--Hubbard model at $U / t = 4$ in Fig.~\ref{fig:ci}. The fixed-particle-number sector contains $\binom{16}{8}^2 = 165{,}636{,}900$ real-space configurations. For the ``Ideal'' curve, we order every configuration by decreasing exact ground-state probability and diagonalize $H$ in nested prefixes of that ordering. This is a favorable probability-ranked reference, not an implementable selection rule. For the ``Sampled'' curve, we instead sample with replacement from the same exact distribution and enlarge the subspace only when a previously unseen configuration appears. We show ten independently sampled trajectories, but they agree on the numbers reported below to better than 0.1\%. Because both constructions use exact ground-state probabilities, the comparison removes state-preparation and hardware errors. At fixed $B$, the separation between the two curves in Fig.~\ref{fig:ci}(a) reflects the stochastic composition of the sampled subspace, whereas Fig.~\ref{fig:ci}(b) charges the sampled trajectory for the samples spent rediscovering configurations already retained. We obtain the CI energies by Lanczos iteration without explicitly forming $H$, using the factorized action
\begin{equation} \label{eq:factorized-action}
  H \Psi = K \Psi + \Psi K^{\mathsf{T}} + U (D \circ \Psi)
\end{equation}
and masking configurations outside the retained subspace \cite{Siro-Harju-2012}.

\begin{figure*}
  \centering
  \begin{minipage}[t]{0.325\textwidth}
    \makebox[\linewidth][l]{\textbf{(a)}}\par
    \includegraphics[width=\linewidth]{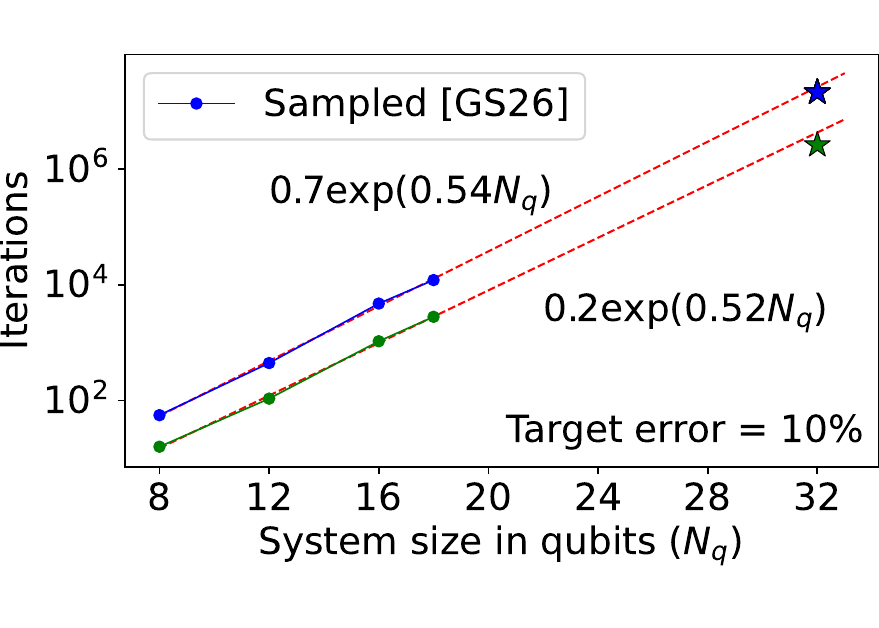}
  \end{minipage}
  \hfill
  \begin{minipage}[t]{0.325\textwidth}
    \makebox[\linewidth][l]{\textbf{(b)}}\par
    \includegraphics[width=\linewidth]{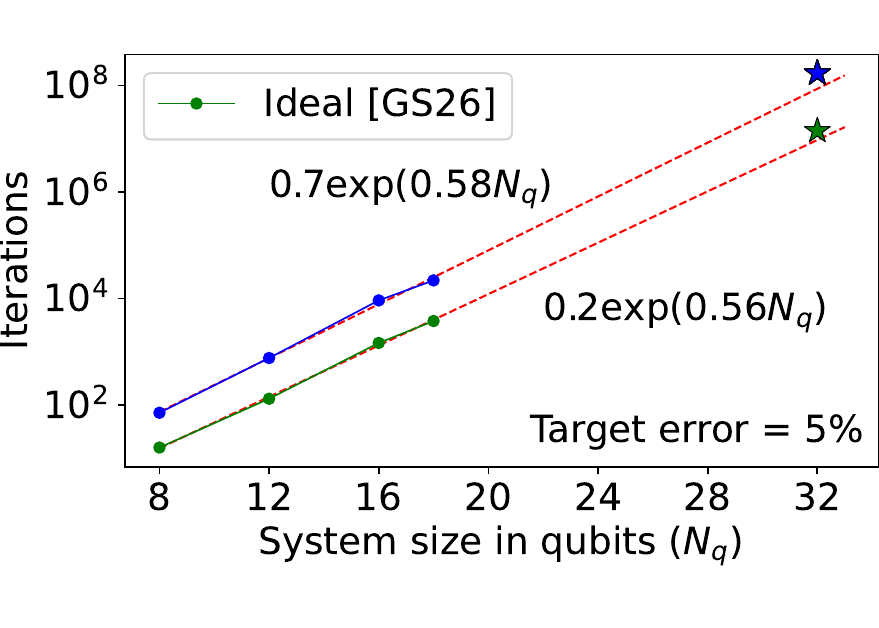}
  \end{minipage}
  \hfill
  \begin{minipage}[t]{0.325\textwidth}
    \makebox[\linewidth][l]{\textbf{(c)}}\par
    \includegraphics[width=\linewidth]{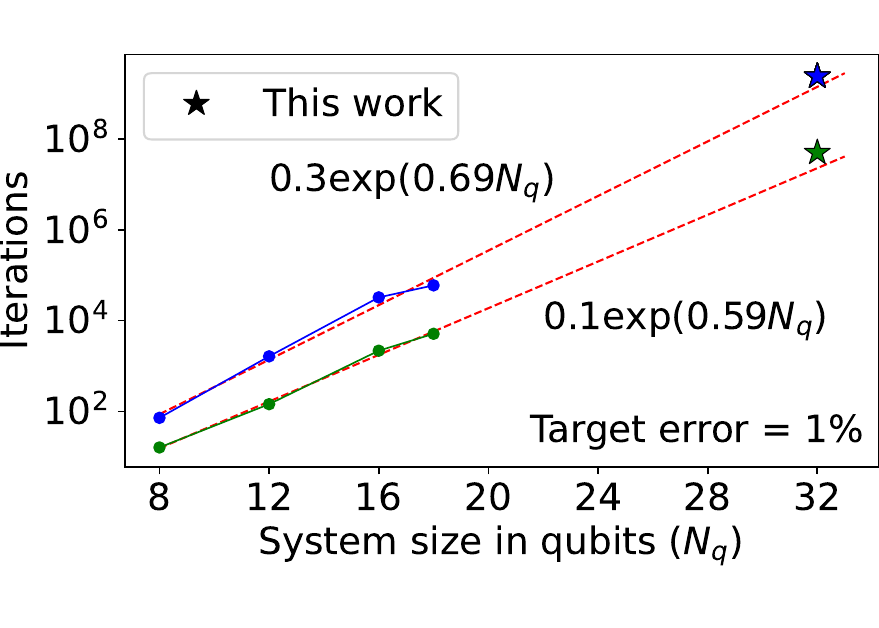}
  \end{minipage}
  \caption{\label{fig:sci-scaling}
    Probability-ranked configuration counts and sampled counts required to reach target relative energy errors of 10\%, 5\%, and 1\% in panels \textbf{(a)}, \textbf{(b)}, and \textbf{(c)}, respectively. The circles reproduce, with permission, the probability-ranked (Ideal) and sampled results of Ref.~\cite{Gaberle-Jattana-2026} for the two-dimensional Fermi--Hubbard model at $U/t=2$ with open boundary conditions. The number of electrons is given in Table II of Ref.~\cite{Cade-Mineh-2020}. The sampled count is the number of samples, whereas the probability-ranked count is the number of retained configurations. Dashed lines are exponential fits to those reference data only. Stars show the corresponding counts for our half-filled, periodic $4 \times 4$ system at $U / t = 4$.
  }
\end{figure*}

Our benchmark follows the raw sampled-configuration subspace of Ref.~\cite{Kanno-Kohda-2026} and the construction called SQD in Ref.~\cite{Gaberle-Jattana-2026}. The SQD implementation of Ref.~\cite{Robledo-Moreno-Motta-2025} instead enlarges the subspace to the outer product of the distinct spin-up and spin-down strings observed. On this $4 \times 4$ system, observing all $\binom{16}{8} = 12{,}870$ strings in each spin sector makes that product the full $1.66 \times 10^8$-dimensional sector and returns the exact energy by construction; observing most of them already produces a nearly full product space. For larger lattices, this saturation is deferred to much larger sample counts: a $6 \times 6$ lattice holds $\binom{36}{18} \simeq 9.1 \times 10^9$ strings per sector, so $10^4$ samples cover a fraction $\sim 10^{-6}$ of one sector rather than most of it. The self-consistent configuration recovery of Ref.~\cite{Robledo-Moreno-Motta-2025} is likewise inapplicable here, since it repairs samples that violate particle-number symmetry and our sampler is exact and noiseless.

Figure~\ref{fig:ci}(a) compares the QSCI convergence with two explicitly correlated benchmarks at equal parameter count, treating each retained configuration as one CI parameter. For the MPS parameter count, we count all symmetry-allowed nonzero tensor entries, excluding structural zeros imposed by the conserved quantum numbers. Table~\ref{tab:feedback-free-comparison} instead excludes the initial MPS and therefore reports only the one and 103 correlator parameters for our method. We use the upper edge of each 95\% bootstrap interval in relative energy error as the crossing criterion, an easier target that again favors QSCI. For the H2-2 interval, the probability-ranked calculation requires $B \simeq 1.7 \times 10^7$ configurations, roughly $7.6 \times 10^4$ times the 225 parameters of the ansatz it is matching. The sampled trajectory crosses at a basis size of $B \simeq 2.5 \times 10^7$ after approximately $2.3 \times 10^8$ samples from the exact ground-state distribution. Thus the number of samples is not set by the final CI dimension: even with exact ground-state sampling, reaching this accuracy requires roughly 9.1 samples per retained configuration.

For the $\chi = 16$ SI interval, the probability-ranked curve reaches its upper edge at $B \simeq 7.6 \times 10^7$, or 46\% of the complete sector, roughly $7.6 \times 10^4$ times the 1012 explicitly correlated parameters. The sampled trajectory reaches the same error at a basis size of $B \simeq 8.6 \times 10^7$, or 52\% of the sector, after approximately $1.06 \times 10^{10}$ samples. The sampled basis is therefore only about 12\% larger than the probability-ranked basis at this crossing, but constructing it requires approximately 124 samples per retained configuration, compared with 9.1 at the H2-2 crossing. Figure~\ref{fig:ci}(b) shows this loss of efficiency directly: the error falls far more slowly against the number of samples than against the basis size, since a growing majority of samples return configurations already retained. Between the two crossings the basis grows by a factor of 3.4 while the samples required grow by a factor of 46.

The large separation from QSCI reflects a structural difference between the two variational ans\"atze. Selected CI enlarges its space one configuration at a time: every configuration outside the subspace has exactly zero amplitude, and the retained coefficients are optimized independently. That hard truncation is inefficient here because the ground-state weight is distributed over a large fraction of the real-space sector, 46\% of it at the $\chi = 16$ crossing above. The correlator instead provides structured parameter sharing: each $\theta_r$ modifies the amplitudes of many configurations simultaneously through occupation patterns that recur throughout the sector. For the positive correlators considered here, this preserves the support and sign structure of the reference. The transcorrelated MPS provides a compact representation of complementary long-range structure in momentum space---909 nonzero tensor entries at $\chi = 16$---despite its weight being spread over many real-space determinants. A comparatively small correlator can therefore correct systematic real-space amplitude correlations rather than resolving the wave function configuration by configuration.

The sample counts separate for a related reason: QSCI spends samples \emph{discovering} basis states. Observing a particular configuration of probability $p$ requires order $1 / p$ samples on average, whereas our estimators spend samples estimating expectation values, whose Monte Carlo error is controlled by estimator variance rather than directly by the size of the support. The advantage is bounded, however. CI converges systematically to the exact energy, whereas the explicitly correlated energy is limited by the variational reach of the reference and correlator.

The system size $L$ of Ref.~\cite{Gaberle-Jattana-2026} denotes the number of qubits; here we write this quantity as $N_q$ to distinguish it from the lattice dimensions. Figure~\ref{fig:sci-scaling} compares our $N_q = 32$ results with the smaller-system scaling data of that reference. The exponential fits in Fig.~\ref{fig:sci-scaling} use only the $N_q = 8$, 12, 16, and 18 qubit data. At a 10\% target error, the sampled fit gives $0.70 \exp(0.544 N_q)$ and predicts $2.59 \times 10^7$ samples at $N_q = 32$, close to the directly calculated $2.08 \times 10^7$. Across all three targets, the probability-ranked and sampled $N_q = 32$ points lie within a factor of 2.2 of the corresponding extrapolations. This agreement is compatible with the exponential trend, but a single additional point cannot establish a common asymptotic exponent, especially because our calculation uses $U / t = 4$, periodic boundaries, and half filling, whereas Ref.~\cite{Gaberle-Jattana-2026} uses $U / t = 2$, open boundaries, and a variety of filling factors. The interaction-strength difference should, if anything, favor QSCI here: increasing $U / t$ at half filling suppresses charge fluctuations, and in the strong-coupling limit the wave function has support only on the $\binom{16}{8} = 12{,}870$ doublon-free configurations.

\subsection{Quantum Krylov subspace diagonalization and unitary variational quantum phase estimation}

\begin{figure}
  \includegraphics[width=\columnwidth]{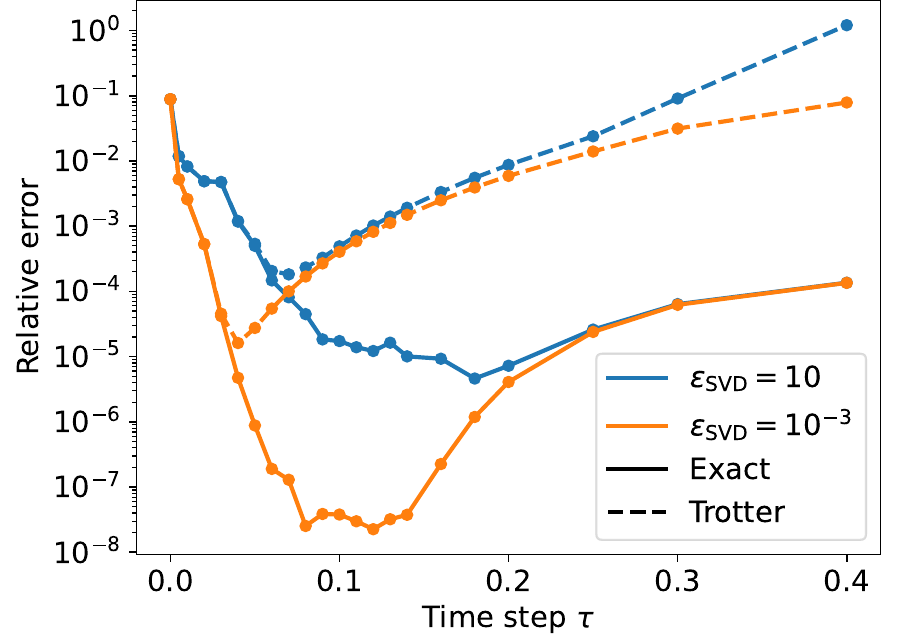}
  \caption{\label{fig:krylov-error-vs-tau}
    QKSD relative energy error with $m = 512$ basis states generated by exact evolution (solid) or second-order Trotter steps (dashed) of size $\tau$. Before solving the GEVP, we discard the singular directions of $S$ with singular values below $\epsilon_\text{SVD}$. The $\tau = 0$ point is the 8.8\% error of the initial state $\ket{\phi_0}$.
  }
\end{figure}

Krylov methods are standard tools for sparse and matrix-free eigenvalue problems, and their quantum analogues inherit that appeal. Unlike our algorithm and QSCI, QKSD and UVQPE improve with circuit depth: lengthening the real-time basis converges toward the exact ground state without changing the prepared state. The cost is repeated real-time evolution, together with phase-sensitive interference requiring either ancilla-controlled evolution or inverse state preparation. Recent demonstrations of Krylov diagonalization on 57 superconducting qubits \cite{Yoshioka-Mezzacapo-2025} and $6 \times 6$ Fermi--Hubbard time evolution on Quantinuum's Helios quantum computer \cite{Granet-Lin-2025} motivate asking what these methods cost on our $4 \times 4$ benchmark under ideal conditions.

All calculations in this section start from the same initial state $\ket{\phi_0}$, a bond-dimension-64 real-space SI MPS with 8.8\% initial energy error and $|\braket{\psi_\text{gs} | \phi_0}|^2 = 0.24$. We expand this MPS in the fixed-particle-number full-CI space and use the factorized action of Eq.~\eqref{eq:factorized-action}. For each base step $\tau$, we generate either the exact propagator $\mathcal{V}(\tau) = e^{-i \tau H}$ or the second-order Trotter step
\begin{equation*}
  \mathcal{V}_\text{T}(\tau) = e^{-i \tau U D / 2} e^{-i \tau T} e^{-i \tau U D / 2}
\end{equation*}
using Krylov-based exponentiation. From the propagated states $\{ \ket{\phi_\ell} = \mathcal{V}(\tau)^\ell \ket{\phi_0} \}_{\ell = 0}^{m - 1}$, we store the autocorrelations $g_\ell = \braket{\phi_0 | \phi_\ell}$ used to compute the Toeplitz overlap matrix $S$ and the UVQPE projected unitary. Exact QKSD also uses $h_\ell = \braket{\phi_0 | H | \phi_\ell}$. Because $H$ does not commute with $\mathcal{V}_\text{T}$, Trotterized QKSD instead retains the propagated vectors and directly computes the dense projected-Hamiltonian data $H_{jk} = \braket{\phi_j | H | \phi_k}$.

Finite sampling perturbs the autocorrelations $g_\ell$ and hence $S$, most severely along its small singular directions. When a total of $M_S$ samples is allocated to the autocorrelations, the analysis of Ref.~\cite{Lee-Lee-Huh-2024} motivates trusting only the singular directions of $S$ whose singular values exceed
\begin{equation} \label{eq:svd-cutoff}
  \epsilon_\text{SVD} = 2 m \sqrt{\frac{2 \ln(2 m)}{M_S}} \ ,
\end{equation}
and discarding the rest; at fixed $m$, $M_S \propto \epsilon_\text{SVD}^{-2}$. We therefore project the matrices onto the left singular vectors of $S$ with singular values strictly greater than $\epsilon_\text{SVD}$ before solving the GEVP; for QKSD we retain the lowest Rayleigh--Ritz energy of the projected pair $(H,S)$.

\begin{figure*}
  \centering
  \vspace{-0.5cm}
  \includegraphics[width=\textwidth]{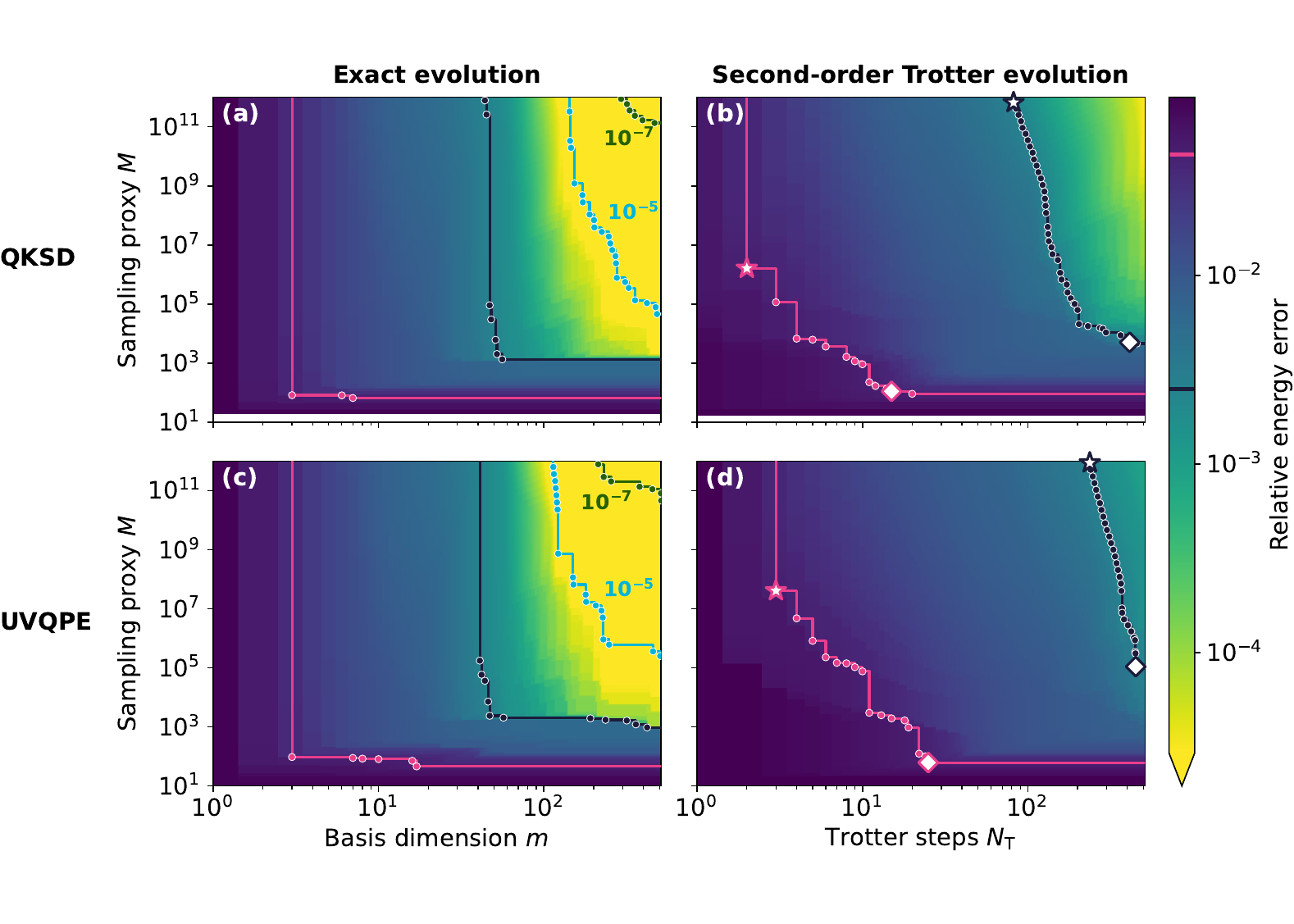}
  \vspace{-1.4cm}
  \caption{\label{fig:krylov-resource-envelopes}
    Resource envelopes for real-time Krylov estimates of the half-filled $4 \times 4$ Fermi--Hubbard ground-state energy at $U / t = 4$, all starting from the same $\chi = 64$ real-space SI MPS. \textbf{(a)} Exact QKSD, \textbf{(b)} second-order-Trotter QKSD, \textbf{(c)} exact UVQPE, and \textbf{(d)} second-order-Trotter UVQPE. Each pixel is the minimum relative energy error over all candidates satisfying the displayed cumulative bounds on the horizontal coordinate (the basis dimension $m$ in the exact panels and the number of sequential Trotter steps $N_\text{T}$ in the Trotter panels) and on $M$, a model-based sampling proxy, not a literal sample count (Eqs.~\eqref{eq:krylov-base-cost} and \eqref{eq:krylov-sampling-costs}); white regions contain no such candidate. The color scale spans the full Trotter-error range; the exact panels saturate and are contoured at $10^{-5}$ and $10^{-7}$. The magenta and navy contours mark the upper edges of the H2-2 (4.40\%) and $\chi = 16$ SI (0.25\%) targets. Stars and diamonds in the Trotter panels mark the minimum-$N_\text{T}$ and minimum-$N_\text{T} M$ operating points in Tables~\ref{tab:feedback-free-comparison} and \ref{tab:krylov-hyperparameters}; the $s = 1$-constrained QKSD points are not marked. Both UVQPE panels restrict $\tau < 0.1061 = 2 \pi / W$.
  }
\end{figure*}

Figure~\ref{fig:krylov-error-vs-tau} compares $\epsilon_\text{SVD} = 10$ and $10^{-3}$, cutoffs whose autocorrelation sampling budgets $M_S$ differ by eight orders of magnitude. At $\tau = 0$ every basis state equals $\ket{\phi_0}$, giving the initial-state energy. Lowering the cutoff enlarges the nested retained Ritz space, so for the exact, mutually consistent projected matrices used here it cannot worsen the QKSD energy; sampled matrices carry no such guarantee \cite{Kirby-2024}. Near $\tau = 0.1$, the tighter cutoff lowers the exact-evolution error by about three orders of magnitude. Bases of closely spaced time-evolved states are necessarily ill-conditioned: as $\tau \to 0$, the basis states become identical, while enlarging the basis at fixed $\tau$ cannot improve the condition number of its Gram matrix. Increasing the time separation therefore provides a route to better conditioning, although the energy improvement in Fig.~\ref{fig:krylov-error-vs-tau} also reflects the longer time interval spanned by the fixed-$m$ basis. Consistent with this picture, the $\epsilon_\text{SVD} = 10$ error decreases up to $\tau \approx 0.18$, and for $\tau \ge 0.25$ resolving the smaller singular directions provides little additional benefit. The Trotterized curves behave differently: beyond $\tau \sim 0.07$, increasing $\tau$ increases the error for both cutoffs. Unlike UVQPE, which projects $\mathcal{V}_\text{T}$ itself, QKSD does not formally require $\mathcal{V}_\text{T}$ to approximate exact evolution, but this deterioration suggests that $\mathcal{V}_\text{T}$ no longer generates a good low-energy subspace. Correspondingly, the benefit of the tighter cutoff is concentrated near $\tau = 0.04$, where it lowers the error by about two orders of magnitude. Trotter error and sampling therefore trade off: an error of $2 \times 10^{-5}$ is reached with exact evolution at $\tau = 0.09$ and $\epsilon_\text{SVD} = 10$, or with Trotterized evolution at $\tau = 0.04$ and $\epsilon_\text{SVD} = 10^{-3}$ at the cost of an increase in $M_S$ by eight orders of magnitude. This motivates separating the base Trotter step from the spacing of the Krylov basis.

By building the basis from every $s$-th propagated state,
\begin{equation*}
  \ket{\varphi_j} = \ket{\phi_{s j}}, \qquad j = 0, \ldots, m - 1 \ ,
\end{equation*}
we can decouple the basis spacing from the Trotter step. For $s > 1$ the basis has a larger effective spacing $s \tau$ while the base Trotter step $\tau$, and hence its Trotter error, remains small. Standard UVQPE has $s = 1$ and $U_{jk} = g_{k - j + 1}$; our generalization still projects the one-base-step propagator, $U_{jk} = g_{s (k - j) + 1}$, so its phase period and aliasing control remain set by $\tau$ (we restrict UVQPE to $\tau < 2 \pi / W = 0.1061$ so the spectrum fits within one phase period). Subsampling offers two separate benefits. First, a larger effective spacing can lift the singular spectrum of $S$ relative to a fixed cutoff, so more directions survive for a given $m$ and $M_S$. Second, for a fixed number of Trotter steps it reduces $m$ by a factor of about $s$, lowering the sampling burden through the $m^2 \ln(2 m)$ dependence obtained by inverting Eq.~\eqref{eq:svd-cutoff}.

To find efficient operating points, we scan candidates $(\tau, s, m, \epsilon_\text{SVD})$ over 
\begin{align*}
  (\tau t) \in \{& 0.005, 0.01, 0.02, 0.03, 0.04, 0.05, 0.06, \\
                 & 0.07, 0.08, 0.09, 0.10, 0.11, 0.12, 0.13, \\
                 & 0.14, 0.16, 0.18, 0.20, 0.25, 0.30, 0.40 \} \ ,
\end{align*}
$s \in [1, 256]$, $m \in [1, 512]$ with $s (m - 1) \leq 512$, and $\epsilon_\text{SVD} = 10^{-6}, 10^{-5.875}, \ldots, 100$ for both exact and Trotterized evolution. All envelopes and operating points below are minima over this scan domain.

For UVQPE, each generalized eigenvalue $\lambda$ of the GEVP $U \V{c} = \lambda S \V{c}$, where $U_{j k} = \braket{\phi_j | \mathcal{V}(\tau) | \phi_k}$ or its Trotterized analogue with $\mathcal{V}_\text{T}(\tau)$, supplies a phase $\vartheta = \arg \lambda$, which we convert to
\begin{equation*}
  E_k = - \frac{\vartheta + 2 \pi k}{\tau},
\end{equation*}
choosing the integer $k$ that places each energy nearest the midpoint $c=(E_\text{min}+E_\text{max})/2$ of the exact spectral interval $[-13.6219,45.6219]$, and then retaining the lowest unwrapped energy. Energies unwrapped outside the interval are retained rather than clipped; the spectral bounds enter only through $c$ and the restriction on $\tau$. This phase selection does not give UVQPE the variational guarantee of Rayleigh--Ritz QKSD.

\begin{table*}
  \caption{\label{tab:krylov-hyperparameters}
    Hyperparameters of the second-order-Trotter QKSD and UVQPE operating points reported in Table~\ref{tab:feedback-free-comparison}, selected from the scan underlying Fig.~\ref{fig:krylov-resource-envelopes}. Here $\tau$ is the base time step, $s$ is the temporal subsampling factor, $m$ is the Krylov dimension, $\epsilon_\text{SVD}$ is the absolute overlap-matrix singular-value cutoff, and $m_\text{SVD}$ is the retained subspace dimension after the SVD truncation. The min.\ $N_\text{T} M$ rows minimize over all scanned $s$, whereas the QKSD min.\ $N_\text{T} M$ ($s = 1$) row fixes $s = 1$. Table~\ref{tab:feedback-free-comparison} gives the resulting $N_\text{T}$ and $M$.
  }
  \begin{ruledtabular}
    \begin{tabular}{llrrrrrrrrrr}
      Method & Operating point    & \multicolumn{5}{c}{H2-2: 4.40\%}                          & \multicolumn{5}{c}{$\chi=16$ SI: 0.25\%} \\ \cline{3-7} \cline{8-12}
             &                    & $\tau$ & $s$ & $m$ & $\epsilon_\text{SVD}$ & $m_\text{SVD}$ & $\tau$ & $s$ & $m$ & $\epsilon_\text{SVD}$ & $m_\text{SVD}$ \\ \hline
      QKSD
             & min. $N_\text{T}$  & 0.16   & 1   & 3   & $1.3 \times 10^{-2}$  & 3              & 0.12   & 1   & 83  & $1.8 \times 10^{-3}$  & 46 \\
             & min. $N_\text{T}M$ & 0.14   & 3   & 6   & 4.2                   & 1              & 0.10   & 7   & 60  & 13                    & 2  \\
             & min. $N_\text{T}M$ ($s=1$) & 0.18 & 1 & 12 & 7.5                 & 1              & 0.13   & 1   & 226 & 24                    & 3  \\
      \hline
      UVQPE
             & min. $N_\text{T}$  & 0.09   & 1   & 3   & $1.8 \times 10^{-3}$  & 3              & 0.05   & 1   & 237 & $1.8 \times 10^{-3}$  & 59 \\
             & min. $N_\text{T}M$ & 0.09   & 1   & 25  & 18                    & 1              & 0.05   & 1   & 449 & 10                    & 7  \\
    \end{tabular}
  \end{ruledtabular}
\end{table*}

We next assign two resources to each solved candidate, representing coherent and sampling costs. For exact evolution the horizontal resource is the basis dimension $m$. For Trotter evolution, it is the number of sequential Trotter steps: $N_\text{T} = s (m - 1) + 1$ for generalized UVQPE and $N_\text{T} = s (m - 1)$ for QKSD. Both horizontal axes extend to 512. The vertical coordinate, $M$, is a model-based sampling proxy, not a literal sample count. Inverting Eq.~\eqref{eq:svd-cutoff} gives the base cost
\begin{equation} \label{eq:krylov-base-cost}
  M_0 = 8 m^2 \ln(2 m) \epsilon_\text{SVD}^{-2} \ .
\end{equation}
Adapting the formulation of Ref.~\cite{Lee-Lee-Huh-2024} to UVQPE and our subsampling procedure, we use
\begin{align} \label{eq:krylov-sampling-costs}
  M = M_0 &
  \begin{cases}
    1 + \norm{H}_\beta & \text{exact QKSD} \\
    1 + \norm{H}_\beta \sqrt{m / 2} & \text{Trotterized QKSD} \\
    f(s) & \text{UVQPE}
  \end{cases} , \\
  f(s) = &
  \begin{cases}
    1 & s = 1 \\
    2 - 1/m & s = 2 \\
    3 - 2/m & s > 2
  \end{cases} \ , \nonumber
\end{align}
where $\norm{H}_\beta$ is the one-norm of the chosen unitary partition. We set $\norm{H}_\beta = 1$, deliberately favorable to QKSD: alternative estimators need not use a full unitary partition---for example, all the interaction and hopping terms can be measured simultaneously in real space and momentum space, respectively \cite{Lee-Choi-2025}---but their fragment variances, covariances, and optimal allocation are not modeled here. The UVQPE factor $f(s)$ counts distinct autocorrelation lags relative to $s = 1$: $m f(s) = m$, $2 m - 1$, and $3 m - 2$ for $s = 1$, $s = 2$, and $s > 2$, respectively.

Letting $\mathcal{N}$ denote the horizontal coordinate, each pixel of Fig.~\ref{fig:krylov-resource-envelopes} is the minimum error among all candidates satisfying both resource bounds,
\begin{equation*}
  \delta_\text{env}(\mathcal{N}, M) = \min_{c: \, \mathcal{N}_c \leq \mathcal{N}, \, M_c \leq M} \delta_c,
\end{equation*}
where candidate $c$ has relative error $\delta_c$ and resources $(\mathcal{N}_c, M_c)$.  We additionally require $M$ to be at least the number of independent real quantities each method must estimate: $2 m f(s)$ for UVQPE, twice the number of distinct autocorrelation lags, $4 m - 3$ for exact QKSD, whose $g_0 = 1$ is known and $h_0$ is real, and $m^2 + 2 m - 2$ for Trotterized QKSD, whose Hermitian $H$ has $m^2$ real entries.

For each target we report two Trotter operating points among target-achieving candidates with $M \leq 10^{12}$: the one with minimum $N_\text{T}$ (``min.\ $N_\text{T}$'') and the one with minimum work proxy $N_\text{T} M$ over all scanned $s$ (``min.\ $N_\text{T} M$''). For QKSD we additionally report the minimum-$N_\text{T} M$ point constrained to $s = 1$ (``min.\ $N_\text{T} M$ ($s = 1$)''), which corresponds to standard QKSD without temporal subsampling and separates the gain from optimizing the ordinary QKSD parameters from the additional gain supplied by subsampling; the UVQPE min.\ $N_\text{T} M$ points already have $s = 1$. Table~\ref{tab:krylov-hyperparameters} lists their parameters.

The exact panels are idealized baselines that charge no evolution time. In this model QKSD and UVQPE perform remarkably similarly: at the minimum-$m M$ points for the $\chi = 16$ target, QKSD requires $m = 56$ and $M \approx 1{,}300$ and UVQPE $m = 47$ and $M \approx 2{,}400$, and both reach $10^{-7}$ within the plotted region. Trotterization hurts UVQPE much more: Trotterized QKSD still reaches errors below $10^{-4}$, whereas Trotterized UVQPE stalls near $10^{-3}$. The gap arises largely because subsampling is free in the exact panels: UVQPE can separate its basis states by $s \tau \gg 2 \pi / W$ while projecting a one-step propagator whose phases remain unaliased. At the exact minimum-$m M$ point, for example, $s \times \tau = 11 \times 0.08$, and Trotterizing this scheme would require 507 steps. QKSD is not subject to this restriction and could reach the same spacing with a larger base step; for it, subsampling accesses effective spacings not present in the finite base-step grid.

In the Trotter panels the min.\ $N_\text{T}$ and min.\ $N_\text{T} M$ points differ sharply. For the $\chi = 16$ benchmark, moving from min.\ $N_\text{T}$ to min.\ $N_\text{T} M$ raises $N_\text{T}$ from 82 to 413 for QKSD and from 237 to 449 for UVQPE, but lowers $M$ by about eight and seven orders of magnitude, respectively; for the H2-2 target, a 7.5--8.3-fold increase in depth lowers $M$ by four to nearly six orders. Most of this reduction does not require subsampling: the QKSD min.\ $N_\text{T} M$ ($s = 1$) point already captures about seven of the eight orders for $\chi = 16$ and 3.9 of 4.2 for H2-2.

\begin{figure}
  \includegraphics[width=\columnwidth]{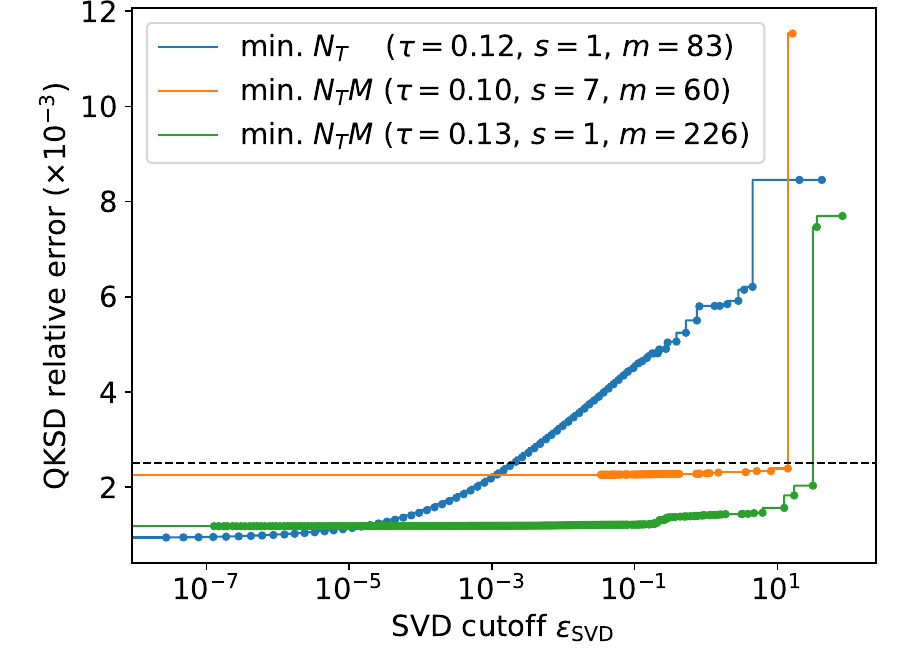}
  \caption{\label{fig:krylov-error-vs-cutoff}
    QKSD relative energy error as a function of the SVD cutoff $\epsilon_\text{SVD}$ for the Krylov bases of the three $\chi = 16$ SI QKSD operating points in Tables~\ref{tab:feedback-free-comparison} and \ref{tab:krylov-hyperparameters}: minimum Trotter depth (min.\ $N_\text{T}$) and minimum work proxy (min.\ $N_\text{T} M$), the latter over all scanned $s$ and constrained to $s = 1$. The $m$ basis states are generated by $\mathcal{V}_\text{T}(\tau)^s \approx \exp(-i s \tau H)$. Markers sit at the singular values of $S$, where the retained subspace gains a direction. Because $M \propto (1 + \sqrt{m/2})\, m^2 \ln(2m)\, \epsilon_\text{SVD}^{-2}$, equal cutoffs do not imply equal $M$: at fixed $\epsilon_\text{SVD}$ the $m = 226$ basis requires about 32 times the $M$ of the $m = 60$ basis. The dashed line marks the 0.25\% target.
  }
\end{figure}

Figure~\ref{fig:krylov-error-vs-cutoff} shows how this reduction arises for the three $\chi = 16$ QKSD points. The min.\ $N_\text{T}$ basis spans a simulated time $\tau N_\text{T}$ of only $9.8 / t$, and its error decreases slowly as singular directions are added, so reaching the target requires 46 directions and $\epsilon_\text{SVD} = 1.8 \times 10^{-3}$. The $s = 1$ min.\ $N_\text{T} M$ basis spans $29 / t$ and is similarly ill-conditioned, with singular values extending to about $10^{-7}$, but its three largest directions already reach the target, so $\epsilon_\text{SVD} = 24$ suffices. Subsampling with $s = 7$ extends the simulated time to $41 / t$ with 60 rather than 226 basis states, and two directions bring the error just below the target, where it plateaus. Relative to the $s = 1$ point, subsampling thus trades 1.8-fold greater coherent depth for a tenfold reduction in $M$ and a 5.6-fold reduction in $N_\text{T} M$, chiefly through the smaller $m$.

Ill-conditioning of closely spaced time-evolved states~\cite{Seki-Yunoki-2021, Lee-Choi-2025, Lee-Lee-Huh-2024} is thus not decisive: it costs samples only when the target requires retaining small singular directions, and the number required here falls sharply as the simulated time grows. These small retained dimensions are, however, empirical properties of the scanned bases. For exact evolution, when the relative phases of distinct energy levels average out over the basis, the nonzero singular values of $S$ are approximately $m P_j$, where $P_j$ is the initial weight of each energy level, including any degeneracy. This suggests why truncation can favor large-overlap components~\cite{Epperly-Lin-2022}. For Trotter evolution, the analogous picture concerns eigenspaces of $\mathcal{V}_\text{T}(\tau)^s$, which can differ from those of $H$. Likewise, the min.\ $N_\text{T} M$ points, with or without subsampling, were found by the scan and are not obviously identifiable a priori.

Two recent time-evolution implementations for 2D Fermi--Hubbard models use Quantinuum's trapped-ion systems. The Jordan--Wigner construction of Ref.~\cite{Alam-Bosse-2025} costs $4 N^{3/2} + 3 N + 5 \sqrt{N}$ compiled two-qubit gates per Trotter step, while the compact encoding of Ref.~\cite{Granet-Lin-2025} uses only $17 N$ at the cost of introducing ancilla qubits. For our $4 \times 4$ system these are 324 and 272 gates per step, respectively. Aside from the corner case of the $4 \times 4$ system with the compact encoding, neither scheme exactly implements $e^{-i \tau T}$; additional approximation of the hopping evolution would introduce a further Trotter error not included here. Exact hopping evolution requires the momentum-space rotation that our algorithm performs only once. For a favorable accounting we use the compact encoding as the point of comparison, disregarding the associated qubit and state-preparation overhead. Within the stated scan and $M \leq 10^{12}$, matching our H2-2 result with QKSD requires two Trotter steps, 544 gates, about $17\%$ more than the 464 of the corresponding SB rotation circuit. Matching the $\chi = 16$ SI benchmark requires 82 steps, 22{,}304 gates, roughly 32 times the 696 of the corresponding SI rotation circuit.

Every modeling choice above favors the Krylov methods. Their reference is a $\chi = 64$ MPS, four times the bond dimension of the $\chi = 16$ result it is asked to match. The sampling model and best-candidate envelopes in this study likewise favor them, while the circuit complexity omits state preparation, phase-sensitive-interference overhead, and hardware noise. The subsampling generalization works in the same favorable direction. Even so, the comparison distinguishes hardware regimes rather than establishing an intrinsic ordering. QKSD and UVQPE retain the stronger long-term route to systematic improvement because a longer real-time basis converges toward the exact ground state; for a fixed reference, our algorithm is limited by the variational family $C(\V{\theta}) \ket{\phi}$, which additional coherent depth does not enlarge.

Systematic improvement with Krylov depth does not imply favorable sampling scaling with system size. The following bound applies to exact-evolution QKSD with arbitrary evolution times under the SVD cutoff and sampling model above. For a target energy $E_\star$ and any threshold $E_+ > E_\star$, define the initial spectral weight at or below the threshold by
\begin{equation*}
  P_+ = \sum_{E_j \leq E_+} |\braket{E_j | \phi_0}|^2.
\end{equation*}

Let $\ket{\psi} = \sum_\ell c_\ell e^{-i t_\ell H} \ket{\phi_0}$ be an exact-evolution QKSD Ritz vector with energy at most $E_\star$. Denote its unnormalized weights at or below $E_+$ and above $E_+$ by $q_-$ and $q_+$, respectively, so that $
\braket{\psi | \psi} = \V{c}^\dagger S \V{c} = q_- + q_+$. For fixed $q_-$ and $q_+$, the most favorable energy distribution places the low-energy weight at $E_\text{gs}$ and the high-energy weight at $E_+$. The target-energy condition therefore requires
\begin{equation*}
  E_\text{gs} q_- + E_+ q_+ \leq \braket{\psi | H | \psi} \leq E_\star (q_- + q_+).
\end{equation*}
Rearranging gives
\begin{equation*}
  q_+ \leq \kappa q_-, \qquad \kappa = \frac{E_\star - E_\text{gs}}{E_+ - E_\star}.
\end{equation*}
Meanwhile, $|\sum_\ell c_\ell e^{-i E_j t_\ell}|^2 \leq m \norm{\V{c}}^2$ limits how much an $m$-dimensional Krylov combination can amplify the initial low-energy weight, giving $q_- \leq m P_+ \norm{\V{c}}^2$. Consequently,
\begin{equation*}
  \V{c}^\dagger S \V{c} \leq (1 + \kappa) m P_+ \norm{\V{c}}^2.
\end{equation*}

Every vector retained by the SVD projection instead satisfies $\V{c}^\dagger S \V{c} > \epsilon_\text{SVD} \norm{\V{c}}^2$. The projected space can therefore contain a Ritz vector reaching $E_\star$ only if $\epsilon_\text{SVD} < (1 + \kappa) m P_+$. Substitution into Eq.~\eqref{eq:krylov-base-cost} gives
\begin{equation*}
  M \geq M_0 > \frac{8 \ln(2 m)}{(1 + \kappa)^2 P_+^2} \ .
\end{equation*}
For the $\chi = 64$ initial state and the $\chi = 16$ target, $|\braket{E_\text{gs} | \phi_0}|^2 = 0.239$, and the first excited level lies $0.146$ above $E_\text{gs}$. Letting $E_+$ approach that level from below gives $\kappa = 0.305$ and $M_0 > 82 \ln(2 m)$. At the minimum-$m M$ exact-evolution QKSD point ($m = 56$), the base cost $M_0 \approx 666$ exceeds this bound by a factor of only 1.72, essentially the $(1 + \kappa)^2 = 1.70$ slack of the bound itself: its cutoff $\epsilon_\text{SVD} = 13.3$ sits just below the phase-averaged ground-state scale $m |\braket{E_\text{gs} | \phi_0}|^2 = 13.4$. The bound is therefore nearly saturated on this benchmark; whether it remains tight at larger sizes is untested, but the scaling argument below requires only that it hold.

For the Fermi--Hubbard model $E_\text{gs}$ is extensive, so for fixed relative errors satisfying $E_0 > E_\star > E_\text{gs}$, $E_0 - E_\star$ and $E_\star - E_\text{gs}$ both grow linearly with $N$, where $E_0 = \braket{\phi_0 | H | \phi_0}$. Choosing $E_+ = (E_0 + E_\star) / 2$ then gives $\kappa = O(1)$, and $P_+$ is the weight of $\ket{\phi_0}$ lying an extensive energy below its mean. With an extensive energy variance, this is $\sim \sqrt{N}$ standard deviations, so a Gaussian energy distribution would give $P_+ = \exp[-\Omega(N)]$. Rigorous concentration bounds exist for spin Hamiltonians with uniformly bounded, finite-range interactions: $P_+ \leq \exp[-\Omega(N)]$ for product states and $P_+ \leq \exp[-\Omega(N^{1/3})]$ for states with exponential clustering of separated observables on a two-dimensional lattice, with clustering constants independent of $N$~\cite{Anshu-2016}. In these settings, the bound requires superpolynomial growth of $M$, regardless of $m$ and the evolution times. Application to the fermionic Hubbard model requires separate justification, and rotated momentum-space MPSs need not satisfy the clustering assumption. For efficiently preparable Hubbard initial states, the scaling of $P_+$ at a fixed energy-density improvement therefore remains open: exponentially decaying ground-state overlaps~\cite{Lee-Lee-Zhai-2023} would not by themselves bound $P_+$, but we know of no such family for which $P_+$ has been shown to decay only polynomially.

\end{document}